\documentclass{emulateapj}
\usepackage{multirow}
\usepackage{amsmath}

\usepackage{color}

\slugcomment{\today}

\newcommand{\mean}[1]{\left \langle #1 \right \rangle}
\newcommand{\kms}{\,km\,s$^{-1}$}
\newcommand{\about}{$\sim\!\!$~}

\newcommand{\lta}{\lesssim}
\newcommand{\gta}{\gtrsim}

\def\arcsec{\hbox{$^{\prime\prime}$}}

\def\farcs{\hbox{$.\!\!^{\prime\prime}$}}
\def\snid{\ifmmode{\rm \tt SNID}\else{\tt SNID}\fi}
\def\dm15{\ifmmode{\Delta m_{15}}\else{$\Delta m_{15}$}\fi}
\def\magarcsec2{\ \rm{mag\ arcsec}^{-2}}
\def\arcsecpixel{\ifmmode{{\rm arcsec~pixel}^{-1}}\else{arcsec~pixel$^{-1}$}\fi}
\def\VR{$V\!R$}
\def\LMC902{SN~2003ma}
\def\BB15K88z{BB15K/88Z}
\shorttitle{The Extremely Energetic \LMC902}
\shortauthors{Rest et~al.}

\begin{document}

\title{Pushing the Boundaries of Conventional Core-Collapse Supernovae: The
Extremely Energetic Supernova \LMC902}

\author{A.~Rest\altaffilmark{1,2}, 
R.~J.~Foley\altaffilmark{3,4}, 
S.~Gezari\altaffilmark{5,6}, 
G.~Narayan\altaffilmark{1}, 
B.~Draine\altaffilmark{7}, 
K.~Olsen\altaffilmark{8,9}, 
M.~E.~Huber\altaffilmark{5,10}, 
T.~Matheson\altaffilmark{9}, 
A.~Garg\altaffilmark{10,11},
D.~L.~Welch\altaffilmark{12},
A.~C.~Becker\altaffilmark{13},
P.~Challis\altaffilmark{3,10},
A.~Clocchiatti\altaffilmark{14}, 
K.~H.~Cook\altaffilmark{10,11},
G.~Damke\altaffilmark{10,15},  
M.~Meixner\altaffilmark{10,16}, 
G.~Miknaitis\altaffilmark{10,17},
D.~Minniti\altaffilmark{14},
L.~Morelli\altaffilmark{18},
S.~Nikolaev\altaffilmark{11},
G.~Pignata\altaffilmark{19},
J.~L.~Prieto\altaffilmark{10,20,21,22},
R.~C.~Smith\altaffilmark{8},
C.~Stubbs\altaffilmark{1},
N.~B.~Suntzeff\altaffilmark{23},
A.~R.~Walker\altaffilmark{8},
W.~M.~Wood-Vasey\altaffilmark{10,24},
A.~Zenteno\altaffilmark{10,25,26},
L.~Wyrzykowski\altaffilmark{27},
A.~Udalski\altaffilmark{28}, 
M.~K.~Szyma{\'n}ski\altaffilmark{28},
M.~Kubiak\altaffilmark{28},
G.~Pietrzy{\'n}ski\altaffilmark{28,29}, 
I.~Soszy{\'n}ski\altaffilmark{28},
O.~Szewczyk\altaffilmark{28,29},
K.~Ulaczyk\altaffilmark{28},
R.~Poleski\altaffilmark{28} 
}


\altaffiltext{1}{Department of Physics, Harvard University, 17 Oxford
Street, Cambridge, MA 02138, email: arest@physics.harvard.edu}
\altaffiltext{2}{Space Telescope Science Institute, 3700 San Martin
Dr., Baltimore, MD 21218}
\altaffiltext{3}{Harvard-Smithsonian Center for Astrophysics, 60 Garden St., 
Cambridge, MA 02138} 
\altaffiltext{4}{Clay Fellow} 
\altaffiltext{5}{Department of Physics and Astronomy, Johns Hopkins University, 3400 N. Charles St., Baltimore, MD 21218} 
\altaffiltext{6}{Hubble Fellow} 
\altaffiltext{7}{Dept.\ of Astrophysical Sciences, Princeton University, Princeton, NJ 08544} 
\altaffiltext{8}{Cerro Tololo Inter-American Observatory, National
Optical Astronomy Observatory (CTIO/NOAO), Colina el Pino S/N, La
Serena, Chile}
\altaffiltext{9}{National Optical Astronomy Observatory, 950 N. Cherry Ave., Tucson, AZ 85719-4933}
\altaffiltext{10}{Visiting Astronomer, Cerro Tololo Inter-American
Observatory, National Optical Astronomy Observatory, which is operated
by the Association of Universities for Research in Astronomy,
Inc. (AURA) under cooperative agreement with the National Science
Foundation}
\altaffiltext{11}{Lawrence Livermore National Laboratory, 7000 East Ave., Livermore, CA 94550}
\altaffiltext{12}{Dept.\ of Physics and Astronomy, McMaster University,
Hamilton, Ontario, L8S 4M1, Canada}
\altaffiltext{13}{Dept.\ of Astronomy, University of Washington, Box 351580, Seattle, WA 98195}
\altaffiltext{14}{Dept.\ of Astronomy, Pontificia Universidad Cat\'olica de Chile, Casilla 306, Santiago 22, Chile}
\altaffiltext{15}{Department of Astronomy, University of Virginia, P.O. Box 400325, Charlottesville, VA 22904-4325}
\altaffiltext{16}{STScI, 3700 San Martin Dr., Baltimore, MD 21218}
\altaffiltext{17}{Center for Neighborhood Technology, 2125 W. North Ave., Chicago IL 60647}
\altaffiltext{18}{Dipartimento di Astronomia, Universit\`{a} di Padova,
vicolo dell'Osservatorio 3, I-35122 Padova, Italy}
\altaffiltext{19}{Departamento de Ciencias Fisicas, Universidad Andres Bello, Avda.
Republica 252, Santiago, Chile}
\altaffiltext{20}{Dept.\ of Astronomy,
Ohio State University, 140 West 18th Ave., Columbus, OH 43210-1173}
\altaffiltext{21}{Carnegie Observatories, 813 Santa Barbara St., Pasadena, CA 91101}
\altaffiltext{22}{Hubble and Carnegie-Princeton fellow}
\altaffiltext{23}{Dept.\ of Physics, Texas A\&M University, College Station, TX 77843-4242}
\altaffiltext{24}{Dept.\ of Physics and Astronomy, University of Pittsburgh, 3951 O'Hara St., Pittsburg, PA 15260}
\altaffiltext{25}{Department of Physics, Ludwig-Maximilians-Universit\"{a}t, Scheinerstr. 1, 81679 M\"{u}nchen, Germany}
\altaffiltext{26}{Excellence Cluster Universe, Boltzmannstr. 2, 85748 Garching, Germany}
\altaffiltext{27}{Institute of Astronomy, University of Cambridge, Madingley~Road, Cambridge~CB3~0HA,~UK}
\altaffiltext{28}{Warsaw University Astronomical Observatory, Al.~Ujazdowskie~4, 00-478~Warszawa, Poland}
\altaffiltext{29}{Universidad de Concepci{\'o}n, Departamento de
Fisica,Astronomy Group, Casilla 160-C, Concepci{\'o}n, Chile}

 
\begin{abstract}
We report the discovery of a supernova (SN) with the highest apparent
energy output to date and conclude that it represents an extreme
example of the Type IIn subclass.  The SN, which was discovered behind
the Large Magellanic Cloud at $z = 0.289$ by the \mbox{SuperMACHO}
microlensing survey, peaked at $M_{R} = -21.5$~mag and only declined
by 2.9~mag over 4.7~years after the peak.
Over this period, \LMC902 had an integrated bolometric luminosity of
$4\times10^{51}$~ergs, more than any other SN to date.  The radiated
energy is close to the limit allowed by conventional core-collapse
explosions.  Optical spectra reveal that \LMC902 has persistent
single-peaked intermediate-width hydrogen lines, a signature of
interaction between the SN and a dense circumstellar medium.  The
light curves show further evidence for circumstellar interaction,
including a long plateau with a shape very similar to the classic
SN~IIn~1988Z -- however, \LMC902 is ten times more luminous at all
epochs.  The fast velocity measured for the intermediate-width
H$\alpha$ component (\about 6000~\kms) points towards an extremely
energetic explosion ($> 10^{52}$~ergs), which imparts a faster
blast-wave speed to the post-shock material and a higher luminosity
from the interaction than is observed in typical SNe~IIn.
Mid-infrared observations of \LMC902 suggest an infrared
light echo is produced by normal interstellar dust at a distance \about
0.5~pc from the SN.
\end{abstract}


\section{Introduction}
\label{sec:intro}

Hydrogen-rich (Type II) core-collapse supernovae (SNe) from massive
stars typically have peak luminosities corresponding to absolute
magnitudes $-14 > M_{R} > -18$~mag that are powered by 
thermal energy that is deposited into the expanding SN
envelope during shock breakout \citep{Li10}.  Depending on the mass of the hydrogen (H)
envelope, a plateau can arise in the optical light curve from a
cooling wave of H recombination that recedes through the ejecta
layers.  After this photospheric phase, the light curve displays an
exponential decline when it is predominantly powered by the
radioactive decay of isotopes created in the explosion ($^{56}$Ni
$\rightarrow ^{56}$Co $\rightarrow ^{56}$Fe).

In the presence of a dense circumstellar medium (CSM), the expanding
SN shock will collide with the CSM and convert the bulk kinetic energy
of the explosion into light \citep{Chevalier94}, and produce a
relatively narrow (2 -- 4 $\times 10^{3}$~\kms) H$\alpha$ line that is
the hallmark of SNe~IIn \citep{Schlegel90}.  The additional
luminosity from the circumstellar interaction can be extremely large
and even dominate the total luminosity.  Furthermore, while the
luminosity from radioactive decay will decline quickly after a few
months, the luminosity from circumstellar interaction can persist at a
near-constant level for several years.

Recently, several extremely luminous SNe have emerged with peak
luminosities with $M_{R} < -21$~mag.  SN~2006gy had a very long rise
time (70~days) and peaked at $M_{R} = -21.8$~mag \citep{Smith07_06gy,
Ofek07}.  SN~2008fz had a similar peak brightness ($M_{\rm V} =
-22.3$~mag) and light-curve shape \citep{Drake10}. SN~2005ap was very
luminous at peak ($M_{\rm unf} = -22.7$~mag), but had a fast rise and
decay.  SN~2008es was similar to SN~2005ap, peaking at $M_{V} =
-22.3$~mag and having a fast rise and decline \citep{Miller09_08es,
Gezari09_08es}.  A self-obscured luminous supernova, SN~2007va, was
detected by \cite{Kozlowski10} only in the infrared with an absolute
mid-IR peak magnitude of $M_{[4.5]} \approx -24.2$.  More recently,
\citet{Quimby09} and \citet{GalYam09} announced the discovery of three
extremely luminous SNe, respectively, which they suggest are powered
by a pulsational pair instability
\citep{Woosley07}. \cite{Pastorello10} finds evidence that these
extremely luminous SNe may be connected to SN~Ic.

A pulsational pair instability is
expected to be important in only the most massive stars -- those
exceeding 95~$M_{\sun}$. The production of electron-positron pairs
results in a contraction and then explosive nuclear burning which
ejects some significant number of solar masses worth of material from
the envelope.  Subsequent repetitions of this sequence of events
result in ejected shells catching up with previous ejected material,
now at much larger radius, producing radiated energy due to the shell
collisions.  This process is estimated to be capable of producing
$10^{50}$~ergs of light and, as importantly, can repeat on short
timescales, providing longer-timescale luminosity.

The energetics of these events push the envelope of our understanding
of stellar evolution.  The peak luminosity, if powered by radioactive
decay, would require \about10~$M_{\sun}$ of $^{56}$Ni.  This very
large amount points to an extremely massive progenitor and the
possibility that these events were the result of a pair-instability SN
\citep{Barkat67, Bond84}.  Alternatively, a significant amount of the
luminosity may be produced by circumstellar interaction, but the mass of the CSM
necessary for the luminosity still indicates that the progenitors must
have been massive stars with significant mass-loss histories.

Furthermore, there is a limit on the energy for which conventional
core-collapse explosions are viable.  Under a conventional
core-collapse explosion, the maximum energy emitted by a SN is
equivalent to the rest mass of a neutron star, a few $10^{54}$~ergs,
with 99\% being emitted as neutrinos \citep[e.g.,][]{Woosley05}.  The
remaining energy, a few $10^{52}$~ergs, is either coupled to the
baryonic material as kinetic energy, or emitted as electromagnetic
radiation.  If a SN has demonstrably greater than a few
$10^{52}$~ergs, the conventional core-collapse scenario must be
re-examined.  This argument has been used for the extremely energetic
broad-lined SNe~Ic associated with gamma-ray bursts
\citep[e.g.,][]{Woosley93, Iwamoto98}; although these constraints are
placed on this class of objects using the kinetic energy, not the
radiated energy.

In the local universe, there have been several well-observed SNe~IIn:
SNe~1988Z \citep{Turatto93}, 1994W \citep{Sollerman98}, 1995N
\citep{Fransson02}, 1998S \citep[e.g.,][]{Leonard00}, 1999el
\citep{DiCarlo02}, 2005ip \citep{Smith09_05ip}, and 2007rt
\citep{Trundle09}.  In addition to their similar spectral evolution,
SNe~IIn all have long plateaus in their light curves after an initial
decline.  This plateau is from shock energy being continuously
converted into visual light through circumstellar interaction.  Once
the shock extends to a radius where the density of the CSM drops, the
source of the luminosity is diminished and the SN fades significantly.

Observations suggest that dust can form in the ejecta of SNe
\citep{Moseley89, Kozasa89, Sugerman06, Smith08_06jc, Rho08, Fox09,
Kotak09_04et}.  This process appears to be enhanced by increasing the
density of the CSM; i.e., SNe~IIn appear to be better at forming dust
than SNe~IIP.  An infrared light echo can appear at late times when
light from the SN explosion reaches a dust shell, and the UV/optical
light is reprocessed into the infrared \citep[e.g.,][]{Gerardy02a}.

Here we present \LMC902, discovered behind the LMC by the
\mbox{SuperMACHO} microlensing survey.  \LMC902 has the light curve
shape and spectral signatures of a SN~IIn, but with a peak luminosity
comparable to the most luminous SNe ever discovered.  In
Section~\ref{sec:observations} we describe the densely-sampled
\mbox{SuperMACHO} and OGLE-III difference imaging photometry (covering
a baseline of 12 years, and the SN light curve over 4.7 years), and
spectroscopy taken at the peak of the SN, and 1, 2, 5, and 6 years later,
and in Section~\ref{sec:comp} we compare the observations to well
studied SNe~IIn.  In Section~\ref{sec:physical} we calculate the total
bolometric luminosity and energy output of the SN, and show that it
exceeds the radiative output of any other SN observed to date.  We
compare possible models for the source of the energy, and show
evidence for an IR echo from variable IR flux measured in archival
{\it Spitzer} IRAC observations.  In Section~\ref{sec:conclusions}, we
summarize our results.

\section{Observations \& Reductions}
\label{sec:observations}

\subsection{Optical Photometry}
\label{sec:optphot}

Starting in 2001, the \mbox{SuperMACHO} Project microlensing survey
used the CTIO 4~m Blanco telescope with its 8K~$\times$~8K MOSAIC
imager (plus its atmospheric dispersion corrector) to monitor the
central portion of the LMC every other night for 5 seasons (September
through December).  The images were taken through our custom ``\VR''
filter ($\lambda_{c} = 625$~nm, $\delta \lambda = 220$~nm; NOAO Code
c6027) with exposure times between 60 and 200 seconds, depending on
the stellar density of each field.  In addition to the \VR\ filter,
images were occasionally obtained through the $B$ and $I$ filters.
Throughout this paper, we denote brightnesses measured in the natural
CTIO magnitude system as $B_{\rm SM}$, \VR$_{\rm SM}$ and $I_{\rm SM}$
as defined in \citet{Miknaitis07}.  To search for variability,
PSF-matched template images were subtracted from search images
\citep{Rest05a, Garg07, Miknaitis07}.  The resulting difference images
are remarkably clean of the (constant) stellar background and are
ideal for searching for variable objects.  Our pipeline detects and
catalogs all variable sources.

\begin{figure}[t]
\begin{center}
\epsscale{1.1}
\plotone{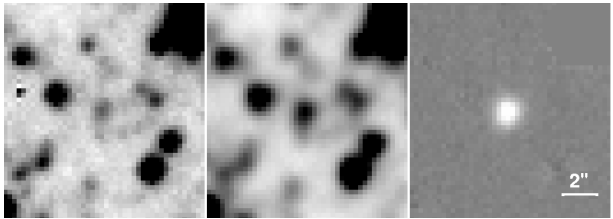}
\caption[]{\mbox{SuperMACHO} images of the region around \LMC902.  Left: Template image from 2001 November 20 (pre-event).  
  Middle: Search image from 2003 December 17,  showing a
  detection of \LMC902 at the center of the image.  Right:
  Difference image (center minus left with an inverted color
  scale) clearly showing the detection of \LMC902.  The white line in
  the lower right corner indicates $2\arcsec$.
  North is up and East is left.

\label{fig:TDpostagestamps}}
\end{center}
\end{figure}

\begin{figure}[t]
\begin{center}
\epsscale{1.5}
\plotone{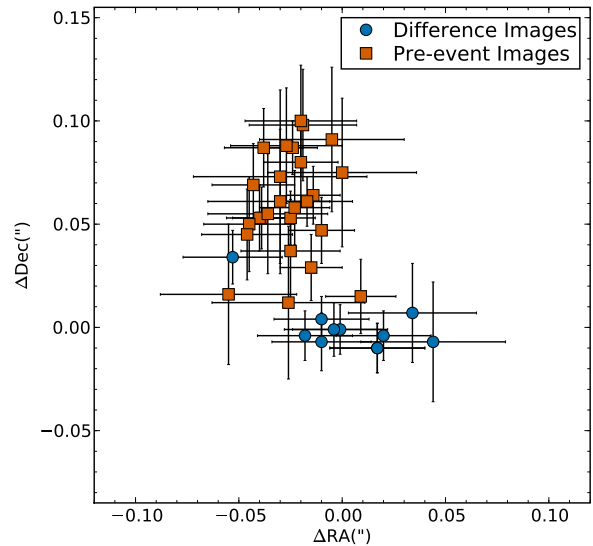}
\caption[]{Centroids of the sources at/near the position of \LMC902
  from multiple images.  
  The red squares indicate the centroids of detections in the
pre-event (MJD $\le 52965$) SuperMACHO images, while the blue circles
indicate the high S/N centroids of \LMC902 in the SuperMACHO
difference images during $52985 \le \mbox{MJD} \le 53020$.  \LMC902 is
offset from the template source by $0.064\pm0.012$~arcsec, 
indicating that it is at a projected distance of $d \approx 460 \pm
85$~pc from the host.
\label{fig:positionplot}}
\end{center}
\end{figure}

On 2003 December 13, the \mbox{SuperMACHO} survey detected a
non-microlensing transient event at position ${\rm RA} =
05$:31:01.878, ${\rm Dec} = -70$:04:15.89 with a \VR~magnitude of 20.53.  
The panels in
Figure~\ref{fig:TDpostagestamps} show from left to right cut-outs from
a pre-event image (2001 November 20), an image 4~days after discovery
(\about 6~days before the peak; 2003 December 17), and their
difference image, respectively.  
The event position in the difference
images is $0.064\pm0.012$~arcsec to the south of a source (the
presumed host galaxy) in the pre-event template image (see
Figure~\ref{fig:positionplot}).  At a redshift of $z = 0.289$ of
\LMC902 and its host (see Section~\ref{ss:spec}), this corresponds to
a projected distance of $d \approx 460 \pm 85$~pc. The host galaxy has
apparent magnitudes of $B_{SM}=20.93 \pm 0.06$~mag, $VR_{SM}=20.15 \pm
0.02$~mag, and $I_{SM}=19.76 \pm 0.03$~mag.

\LMC902 was also detected by the OGLE--III survey (2001 -- 2009), which
is a photometric survey using a dedicated 1.3-m Warsaw telescope
located at the Las Campanas Observatory, Chile, operated by the
Carnegie Institution of Washington.  The camera uses eight SITe
$2048\times4096$ CCD detectors with 15~\micron\ pixels resulting in
0.26~\arcsecpixel\ scale and $35\times35$~arcmins total field of
view. For the details of the instrumentation setup we refer the reader
to \citet{Udalski03}.  Approximately 500 photometric points per star
towards the LMC were accumulated over eight seasons, between July~2001
and May~2009. Most of the observations were carried through the
Cousins {\it I}-band filter with exposure times 180~seconds.  Full
description of the reduction techniques, difference image analysis,
photometric calibration and astrometric transformations can be found
in \citet{Wozniak00, Udalski08}. 

Pre-event {\it I}-band data were also gathered during the OGLE--II
survey (1996 -- 2000), which used the same dedicated telescope as
OGLE--III, but was equipped with a single $2048\times2048$ CCD
\citep{Udalski97}. Since the QE of the two CCD's are slightly
different, we apply a small 2\% correction to the OGLE--II data so
that their average flux matches the pre-event flux in the OGLE--III.
Since the OGLE-II data is several years before the event, we used it
to test for pre-event variability, for example caused by an active
galactic nucleus (AGN).

\begin{figure}[t]
\begin{center}
\epsscale{1.18}
\plotone{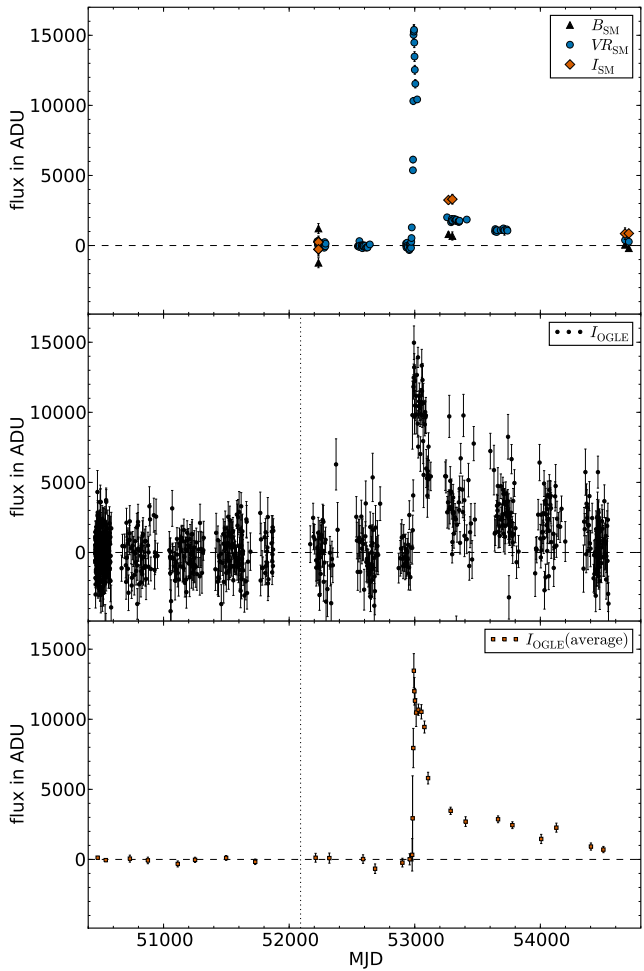}
\caption[]{Difference image light curves of \LMC902. Top:
  Black triangles, blue circles, and orange diamonds show the
  \mbox{SuperMACHO} difference flux light curves in $B_{\rm SM}$, \VR$_{\rm
  SM}$, and $I_{\rm SM}$, respectively, from 2001 November 11  to
  2008 August 25.  Middle: Small black circles indicate
  the $I_{\rm OGLE}$ difference flux light curves.  Bottom: Because of the low
  S/N of a single detection, we average the data (orange squares). 
  The horizontal dashed lines indicate zero flux. 
  The vertical dotted line 
  indicates the transition between OGLE--II and OGLE--III measurements.
\label{fig:TDflux}}
\end{center}
\end{figure}

The upper and lower panels of Figure~\ref{fig:TDflux} shows the
difference image light curves from the \mbox{SuperMACHO} and OGLE
surveys. Images with poor seeing and/or significant cloud cover are
more likely to have large (and non-Poissonian) systematic
uncertainties. Therefore, only images with seeing better than
$1.5\arcsec$ and with atmospheric extinction less than 1~mag are used
for the analysis. We make an exception for images from July and
August~2008, which typically have poor seeing, but are crucial for
understanding the late-time behavior of \LMC902.  For these months,
images with seeing better than $1.9\arcsec$ were retained.  Since
there are several images in all filters from 2008, we were able to
average the flux from these images.

Because difference imaging reveals more instrumental and reduction
artifacts than there are real variable objects in an image, standard
profile-fitting software package, like DoPHOT \citep{Schechter93},
have problems determining the optimal PSF in the difference image.  To
remedy this, we implement a customized version of DoPHOT that uses the
applicable PSF from the original image \citep{Rest05a, Garg07}.  After
an object is detected in any difference image we ``force'' photometry
in all difference images at the event position regardless of the flux
level.  We thus measure the flux in pre-event images and negative
fluxes are possible.  These measurements are critical for constraining
the light curves.  For example the pre-event light curve (${\rm MJD} <
52970$) is consistent with no variability, indicating that the host
galaxy has not demonstrated a history of variability that would
indicate the presence of an AGN.

In addition to the \mbox{SuperMACHO} \VR\ images, several $B$ and $I$
images were obtained on 2001 November 20, 2004 September 20, 2004
October 21, 2008 July 27, and 2008 August 26.  The images from 2001
November 20, obtained before the event, are used as the template
images for these filters.

The middle panel of Figure~\ref{fig:TDflux} shows the OGLE $I$-band
difference image photometry (small black circles).  Mainly because of
the difference in telescope aperture (1.3~m for OGLE and 4~m for
\mbox{SuperMACHO}), the signal-to-noise ratio (S/N) of the OGLE photometry is
significantly lower than that of the \mbox{SuperMACHO} data.  
To obtain measurements with sufficient S/Ns, we have binned the OGLE
data (lower panel of Figure~\ref{fig:TDflux}). Since the OGLE
observations are taken in blocks, we divide each block in two equal
parts, each containing the same amount of measurements. The only
exception is during peak, where the S/N of the measurements is
significantly higher, and in addition the light curve evolves
faster. Thus around peak ($52975<=\mbox{MJD}<=53015$), we average the
data in 5-day bins, and in 25-day bins in the range
$53015<\mbox{MJD}<=53150$. In the following analysis, we only use the
averaged OGLE data.

%

\begin{figure}[t]
\begin{center}
\epsscale{1.18}
\plotone{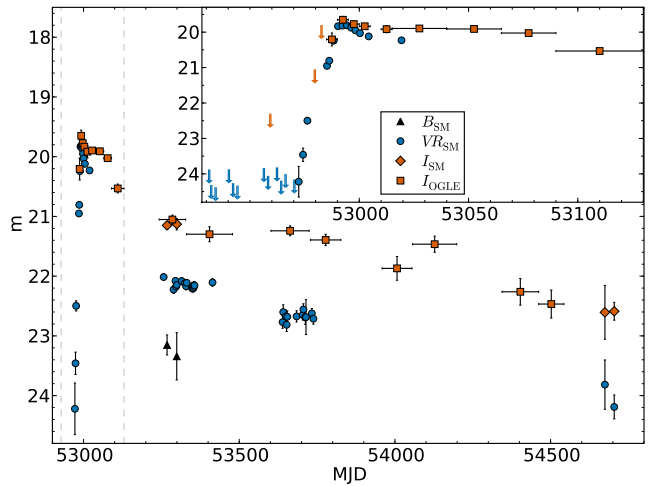}
\caption[]{Optical light curves of \LMC902.  The black triangles, blue
circles, orange diamonds, and orange squares show the apparent
magnitudes in the natural system of $B_{\rm SM}$, \VR$_{\rm SM}$,
$I_{\rm SM}$, and $I_{\rm OGLE}$, respectively.  Because of the low
S/N of a single detection, we average $I_{\rm OGLE}$ (orange squares),
with the range of epochs used to calculate this average indicated by
the error bars along the time axis.  The inset shows the light curve
during peak from $52928 \le {\rm MJD} \le 53130$, 
and the $3\sigma$
brightness upper limits.
\label{fig:apparentmags}}
\end{center}
\end{figure}

We convert the CTIO 4~m difference flux detections into apparent
magnitudes (see Figure~\ref{fig:apparentmags} and
Table~\ref{tab:apparentmags}) in the natural CTIO 4~m magnitude
system, and we correct for extinction as described in
Section~\ref{sec:extinction}. The OGLE measurements have been
converted into standard I-band magnitudes using the standard OGLE
photometric calibration \citep{Wozniak00, Udalski08}.

\begin{deluxetable*}{ccccc}
\tablewidth{0pt}
\tablecaption{Apparent magnitudes of \LMC902\label{tab:apparentmags}}
\tabletypesize{\tiny}
\tablehead{
\colhead{MJD} & \colhead{$B_{\rm SM}$} & \colhead{\VR$_{\rm SM}$} & \colhead{$I_{\rm SM}$} & \colhead{$I_{\rm OGLE}$}
}
\startdata
 52959.29126 &        \nodata &        \nodata &        \nodata & $<22.30$ \\
 52964.27804 &        \nodata & $<24.19$       &        \nodata &        \nodata\\
 52966.28245 &        \nodata & $<24.01$       &        \nodata &        \nodata\\
 52970.30195 &        \nodata & $<24.16$       &        \nodata &        \nodata\\
 52972.29799 &        \nodata & $24.22$ (0.43) &        \nodata &        \nodata\\
 52974.25935 &        \nodata & $23.46$ (0.19) &        \nodata &        \nodata\\
 52976.26257 &        \nodata & $22.50$ (0.09) &        \nodata &        \nodata\\
 52979.64149 &        \nodata &        \nodata &        \nodata & $<21.05$ \\
 52982.71623 &        \nodata &        \nodata &        \nodata & $<19.80$ \\
 52985.24459 &        \nodata & $20.95$ (0.03) &        \nodata &        \nodata\\
 52986.27416 &        \nodata & $20.81$ (0.03) &        \nodata &        \nodata\\
 52988.68744 &        \nodata &        \nodata &        \nodata & $20.21$ (0.18)\\
 52988.28817 &        \nodata & $20.24$ (0.02) &        \nodata &        \nodata\\
 52990.25875 &        \nodata & $19.83$ (0.02) &        \nodata &        \nodata\\
 52992.29069 &        \nodata & $19.82$ (0.03) &        \nodata &        \nodata\\
 52992.73489 &        \nodata &        \nodata &        \nodata & $19.65$ (0.10)\\
 52994.31115 &        \nodata & $19.81$ (0.02) &        \nodata &        \nodata\\
 52996.28857 &        \nodata & $19.87$ (0.02) &        \nodata &        \nodata\\
 52997.44242 &        \nodata &        \nodata &        \nodata & $19.77$ (0.09)\\
 52998.25951 &        \nodata & $19.95$ (0.02) &        \nodata &        \nodata\\
 53000.31187 &        \nodata & $20.03$ (0.02) &        \nodata &        \nodata\\
 53001.72660 &        \nodata &        \nodata &        \nodata & $19.83$ (0.08)\\
 53004.32118 &        \nodata & $20.12$ (0.02) &        \nodata &        \nodata\\
 53011.28573 &        \nodata &        \nodata &        \nodata & $19.92$ (0.10)\\
 53019.31815 &        \nodata & $20.23$ (0.02) &        \nodata &        \nodata\\
 53029.29812 &        \nodata &        \nodata &        \nodata & $19.90$ (0.04)\\
 53052.97253 &        \nodata &        \nodata &        \nodata & $19.91$ (0.05)\\
 53077.45696 &        \nodata &        \nodata &        \nodata & $20.02$ (0.05)\\
 53106.03597 &        \nodata &        \nodata &        \nodata & $20.53$ (0.07)\\
 53257.38920 &        \nodata & $22.02$ (0.05) &        \nodata &        \nodata\\
 53268.26639 & $23.15$ (0.17) &        \nodata &        \nodata &        \nodata\\
 53268.27047 &        \nodata &        \nodata & $21.15$ (0.06) &        \nodata\\
 53285.33774 &        \nodata &        \nodata &        \nodata & $21.05$ (0.07)\\
 53289.21636 &        \nodata & $22.22$ (0.06) &        \nodata &        \nodata\\
 53295.33515 &        \nodata & $22.08$ (0.06) &        \nodata &        \nodata\\
 53297.22518 &        \nodata & $22.18$ (0.06) &        \nodata &        \nodata\\
 53299.36194 &        \nodata & $22.15$ (0.06) &        \nodata &        \nodata\\
 53299.36544 & $23.34$ (0.40) &        \nodata &        \nodata &        \nodata\\
 53299.36742 &        \nodata &        \nodata & $21.13$ (0.10) &        \nodata\\
 53315.28817 &        \nodata & $22.08$ (0.06) &        \nodata &        \nodata\\
 53325.24440 &        \nodata & $22.12$ (0.06) &        \nodata &        \nodata\\
 53329.29664 &        \nodata & $22.17$ (0.06) &        \nodata &        \nodata\\
 53331.34066 &        \nodata & $22.11$ (0.06) &        \nodata &        \nodata\\
 53348.18819 &        \nodata & $22.20$ (0.06) &        \nodata &        \nodata\\
 53350.32545 &        \nodata & $22.17$ (0.07) &        \nodata &        \nodata\\
 53352.23171 &        \nodata & $22.21$ (0.06) &        \nodata &        \nodata\\
 53354.23758 &        \nodata & $22.18$ (0.06) &        \nodata &        \nodata\\
 53356.25495 &        \nodata & $22.16$ (0.07) &        \nodata &        \nodata\\
 53404.63460 &        \nodata &        \nodata &        \nodata & $21.30$ (0.12)\\
 53414.22055 &        \nodata & $22.11$ (0.06) &        \nodata &        \nodata\\
 53639.30253 &        \nodata & $22.77$ (0.10) &        \nodata &        \nodata\\
 53641.33128 &        \nodata & $22.60$ (0.12) &        \nodata &        \nodata\\
 53649.38498 &        \nodata & $22.68$ (0.11) &        \nodata &        \nodata\\
 53652.32054 &        \nodata & $22.81$ (0.11) &        \nodata &        \nodata\\
 53654.22569 &        \nodata & $22.68$ (0.10) &        \nodata &        \nodata\\
 53663.32678 &        \nodata &        \nodata &        \nodata & $21.24$ (0.08)\\
 53684.22316 &        \nodata & $22.67$ (0.09) &        \nodata &        \nodata\\
 53704.30592 &        \nodata & $22.65$ (0.09) &        \nodata &        \nodata\\
 53706.22709 &        \nodata & $22.56$ (0.10) &        \nodata &        \nodata\\
 53712.35217 &        \nodata & $22.69$ (0.12) &        \nodata &        \nodata\\
 53714.35882 &        \nodata & $22.69$ (0.29) &        \nodata &        \nodata\\
 53733.16559 &        \nodata & $22.62$ (0.08) &        \nodata &        \nodata\\
 53738.16874 &        \nodata & $22.71$ (0.09) &        \nodata &        \nodata\\
 53777.61556 &        \nodata &        \nodata &        \nodata & $21.39$ (0.10)\\
 54006.82171 &        \nodata &        \nodata &        \nodata & $21.87$ (0.20)\\
 54127.67356 &        \nodata &        \nodata &        \nodata & $21.46$ (0.14)\\
 54402.73904 &        \nodata &        \nodata &        \nodata & $22.26$ (0.22)\\
 54502.13612 &        \nodata &        \nodata &        \nodata & $22.47$ (0.23)\\
 54674.41064 &        \nodata & $23.82$ (0.41) &        \nodata &        \nodata\\
 54674.42468 &        \nodata &        \nodata & $22.61$ (0.45) &        \nodata\\
 54704.35730 &        \nodata & $24.19$ (0.20) &        \nodata &        \nodata\\
 54704.37563 &        \nodata &        \nodata & $22.59$ (0.15) &        \nodata
\enddata
\end{deluxetable*}

\subsection{Infrared Photometry\label{sec:IRredux}}

We use archived images and catalogs from the {\it Spitzer} Space Telescope
{\it SAGE} survey of the LMC \citep{Meixner06} to measure the
mid-infrared photometry of \LMC902.  The {\it SAGE} survey covered a
$7^{\circ} \times 7^{\circ}$ region centered on the LMC during two
epochs, 2005 July 15 -- 26 (epoch 1) and 2005 October 26 -- November 2 (epoch
2) for its IRAC observations, and 2005 July 27 -- August 3 (epoch 1) and
2005 November 2 -- 9 (epoch 2) for its MIPS observations.  We used the
epoch 1 and 2 images provided by the {\it SAGE} team to produce small
(300~pixel wide) cutouts around the position of \LMC902, for all four
IRAC bands and the 24~\micron\ MIPS band.  The IRAC and MIPS cutouts
have spatial scales of 1.2 and 2.5~\arcsecpixel, respectively,
and are all on a common coordinate reference frame.

Although an object was clearly visible at the position of \LMC902
on several of the {\it Spitzer}
images, we found that the {\it SAGE} catalog did not contain
photometry for it in all bands, likely due to the presence of several
bright stars near the \LMC902 position.  Moreover, because the
crowding in the {\it Spitzer} images forced us to conduct artificial star
tests in order to understand the true photometric errors, we performed
our own PSF-fitting photometry on the image cutouts using
DAOPHOT/ALLSTAR \citep{Stetson87}.  We used a threshold of 
1$\sigma$ to find stars and a sequence of apertures for initial
photometry.  To derive a PSF for each image, we first selected 30 -- 130
PSF stars for the IRAC images, and 18 for the MIPS 24~\micron\ image.
We then used an iterative procedure of fitting a PSF function to the
PSF stars, subtracting neighboring sources, and then refitting the PSF
function to the PSF stars, while increasing the order of the spatial
variability of the function up to a maximum of two.  After several
iterations, we used the derived PSF to derive photometry for all detected
sources. After subtracting all sources except for the PSF stars,
we calculated aperture photometry of the PSF
stars out to a radius of 11~pixels (13$\farcs$2) for the IRAC images,
and 29~pixels (72$\farcs$5) for the MIPS 24~\micron\ images.  We used
this aperture photometry to apply a correction to the PSF-fitted
photometry.  Finally, we matched the aperture-corrected photometry to 
that of sources in the {\it SAGE} catalog, and applied the zero-point 
differences to provide the photometric calibration.

For three of the images (epochs 1 and 2 for the 3.6~\micron\ band, and
epoch 1 for the 4.5~\micron\ band), 
we found that the crowding in the
images and the intrinsic faintness of the object left \LMC902 still
unmeasured, with no matching sources within 3 pixels of its position.
For these three images, we thus use the 8.0~\micron\ epoch 1 image
positions as the master source list, and forced ALLSTAR to fit
photometry at the positions, allowing only the sky and object
magnitude to be free parameters.

To ensure that our derived photometric uncertainties were reasonable in
the {\it Spitzer} images which all suffer from crowding, we performed
artificial star tests. Artificial stars were added, a few at a time, to
copies of the original images, measured using the photometric
procedure described above, and then analyzed to evaluate the
photometric errors, offsets, and completeness as a function of input
magnitude. In the course of these tests, we added a total of 10,000
artificial stars, with magnitudes distributed evenly over the range
$14 \le M_{3.6~\mu\rm{ m}} \le 18$~mag, and with colors with respect to
the other bands matched to those observed for \LMC902.  The stars were
added 50 at a time (\about 5\% of the real detected sources in each
image) at random positions.  Following the photometric measurement
step, we recovered the detected artificial stars by first removing
from the output lists all stars matching real sources detected in the
original images, and then comparing the remaining sources with the
list of input artificial stars.  We then computed the average offset
and dispersion in the input minus recovered magnitudes and the
fraction of artificial stars recovered as a function of input
magnitude.  

Table~\ref{tab:spitzer} lists the {\it Spitzer} photometry of \LMC902,
and the quantities derived from the artificial star tests.  The
measured flux in mJy is shown in the 3rd column. The flux uncertainty
in the 4th column was determined by DAOPHOT and is typically between
5\% - 15\%. This is the appropriate relative uncertainty between the
fluxes measured in the same band but different epochs, 
since any bias due to crowding should have nearly the same effect in both epochs assuming that the
neighboring objects are not variable. For different wavelengths, the
crowding is different and thus has a different impact on the
photometry. Therefore our estimate for the total photometric
uncertainty (6th column in Table~\ref{tab:spitzer}) of \LMC902 at each
epoch is based on the mean photometric uncertainty arrived at in the
appropriate magnitude range of the artificial stars lists.  For all
comparisons of fluxes from different bands, we use these total errors,
which are \about 30\%, \about 30\%, \about 30 -- 50\%, \about 10 --
15\%, and \about 20\% for the 3.6, 4.5, 5.8, 8.0, and 24~\micron\
images, respectively.




The average absolute offsets in the input minus recovered magnitudes
of the artificial stars are typically smaller than 0.1 mags, such that
we do not expect the Spitzer detections of SN2003ma to be spurious.
The only exception is the 5.8 micron photometry, for which the
absolute offsets are ~0.5 mags.  The average completeness for the
5.8~\micron\ images at the magnitude of \LMC902 was also $<$50\%,
which suggests that the 5.8~\micron\ photometry is unreliable compared
to that in the other bands. As a consistency check, we compared our
unforced photometry to the recently complete, final catalog and
archive lists from the {\it SAGE} project \citep{Meixner06} and found
them to be in agreement.
\begin{deluxetable*}{cccccccc} 
\tablewidth{0pt}
\setlength{\tabcolsep}{1in} 
\tabletypesize{\scriptsize}
\tablecaption{IRAC and MIPS Photometry and Artificial Star Test Results\label{tab:spitzer}}
\tablehead{
\colhead{Band}       & \colhead{t\tablenotemark{a}} & \colhead{Flux\tablenotemark{b}} & \colhead{Flux relerr\tablenotemark{c}} & \colhead{$\Delta$Flux\tablenotemark{d}} & \colhead{Flux err\tablenotemark{e}} & \colhead{Compl.\tablenotemark{f}}\\
\colhead{(\micron)}  &                                  & \colhead{(mJy)}& \colhead{(mJy)}                       & \colhead{(mJy)}                           & \colhead{(mJy)}                         & \colhead{(\%)}
}
\startdata
\phantom{2}3.6 &      1  &     0.063\tablenotemark{g}           & 0.006 & 0.006 &         0.018 &             46.4\\
\phantom{2}4.5 &      1  &     0.048\tablenotemark{g}           & 0.003 & 0.003 &         0.014 &             53.3\\
\phantom{2}5.8 &      1  &     0.059\phantom{\tablenotemark{g}} & 0.004 & 0.017 &         0.017 &             38.9\\
\phantom{2}8.0 &      1  &     0.108\phantom{\tablenotemark{g}} & 0.004 & 0.014 &         0.018 &             77.3\\
          24.0 &     1  &     1.164\phantom{\tablenotemark{g}} & 0.080 & 0.099 &         0.227 &             94.4\\
\hline
\phantom{2}3.6 &      2  &     0.055\tablenotemark{g}           & 0.003 & 0.005 &         0.019 &             40.9\\
\phantom{2}4.5 &      2  &     0.065\phantom{\tablenotemark{g}} & 0.003 & 0.006 &         0.014 &             65.6\\
\phantom{2}5.8 &      2  &     0.028\phantom{\tablenotemark{g}} & 0.004 & 0.015 &         0.013 &             13.6\\
\phantom{2}8.0 &      2  &     0.184\phantom{\tablenotemark{g}} & 0.006 & 0.014 &         0.022 &             89.3\\
          24.0 &     2  &     1.141\phantom{\tablenotemark{g}} & 0.108 & 0.098 &         0.228 &             94.0
\enddata
\tablenotetext{a}{Epoch ID. Epoch 1 spans from 2005 July 15 -- August 3, and epoch 2 from 2005 October 26 -- November 9.}
\tablenotetext{b}{
Flux at the position of \LMC902. This flux contains both the flux of \LMC902 as well as any potential flux from the host galaxy.}
\tablenotetext{c}{Flux error as determined by DAOPHOT. This is the appropriate relative error between the fluxes measured in the same band at different epochs
} 
\tablenotetext{d}{Average detected flux minus input flux as derived from artificial star tests, interpolated at the measured flux level of \LMC902 in each image.}
\tablenotetext{e}{Average dispersion in the detected flux minus the input flux from the artificial star tests, interpolated at the measured flux level of \LMC902 in each image.}
\tablenotetext{f}{Average completeness in \%.}
\tablenotetext{g}{Forced photometry.}
\end{deluxetable*}

\subsection{Spectroscopy}\label{ss:spec}

\begin{figure}[t]
\begin{center}
\epsscale{0.9}
\rotatebox{90}{
\plotone{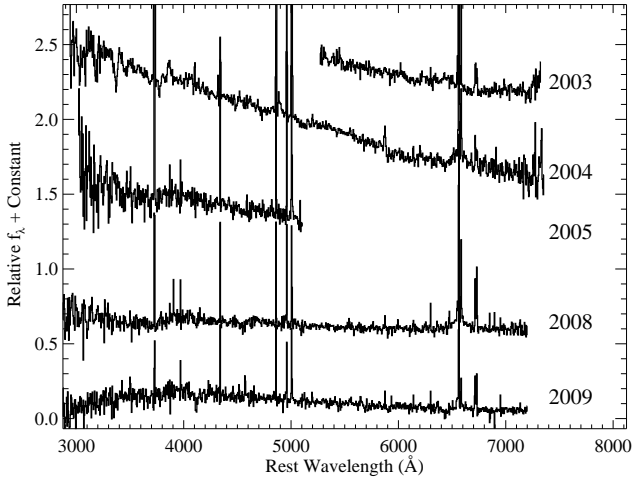}}
\caption[]{Optical spectra of \LMC902.  The spectra are denoted by the
  year in which they were obtained.  The spectra in years 2003, 2004,
  2005, and 2008 correspond to rest-frame phases of 1, 279, 905, and
  1423~days relative to $B$ maximum, respectively.  There is
  significant galaxy contamination in all spectra, likely dominating
  the spectrum at later phases.  Since the 2004 spectrum was not
  obtained at the parallactic angle, its spectral shape is likely
  incorrect.
\label{fig:spec}}
\end{center}
\end{figure}

We followed \LMC902 spectroscopically for nearly six years using
the 8~m Gemini-South telescope (with GMOS; \citealt{Hook04}) and the
6.5~m Magellan Clay telescope (with LDSS2;
\citealt{Allington-Smith94},
LDSS3\footnote{http://www.lco.cl/telescopes-information/magellan/instruments-1/ldss-3-1/
.}, and MagE; \citealt{Marshall08}).  The spectra\footnote{The Gemini
South GMOS spectra were obtained as part of the GS-2003B-Q-12 science
program and the header target name in the images and spectra was
cand10194.sm76\_4.} are presented in Figure~\ref{fig:spec} and a
journal of observations are presented in Table~\ref{tab:spec}.

\begin{deluxetable}{lllr}
\tabletypesize{\scriptsize}
\tablewidth{0pt}
\tablecaption{Log of Spectral Observations\label{tab:spec}}
\tablehead{
\colhead{Phase\tablenotemark{a}} &
\colhead{UT Date} &
\colhead{Telescope /} &
\colhead{Exposure} \\
\colhead{} &
\colhead{} &
\colhead{Instrument} &
\colhead{(s)}}

\startdata

   1 & 2003 December 22.3 & Gemini South/GMOS & $3 \times 610$  \\
 279 & 2004 December 14.1 & Magellan/LDSS2    & 1500            \\
 905 & 2005 October  29.3 & Magellan/LDSS3    & 900             \\
1423 & 2008 December 28.3 & Magellan/MagE     & $2 \times 1800$ \\
1649 & 2009 October  17.3 & Magellan/MagE     & 1800, 1200

\enddata

\tablenotetext{a}{Rest-frame days since $B$ maximum, 2003 December 20.8 (JD 2,452,994.3).}

\end{deluxetable}

Standard CCD processing and spectrum extraction were accomplished with
IRAF.  The data were extracted using the optimal algorithm of
\citet{Horne86}.  Low-order polynomial fits to calibration-lamp
spectra were used to establish the wavelength scale, and small
adjustments derived from night-sky lines in the object frames were
applied.  For the MagE spectrum, the sky was subtracted from the images
using the method described by \citet{Kelson03}.  We employed our own
IDL routines to flux calibrate the data and remove telluric lines
using the well-exposed continua of the spectrophotometric standards
\citep{Wade88, Foley03, Foley09:08ha}.

The 2004 spectrum was obtained far from the parallactic angle at high
airmass.  As a result, its continuum and any derived line ratios are
likely incorrect \citep{Filippenko82}.  Since the offset between the
SN and host is very small, we were not able to perform local galaxy
subtraction during extraction.  For the 2003 and 2008 spectra, we have
calibrated our absolute spectrophotometry to match concurrent
photometry of the SN and host combined.  This is not possible for the
2004 and 2005 spectra because of incorrect relative spectrophotometry
and a small wavelength range that does not fully cover any of our
broad-band filters, respectively.

\begin{figure}[t]
\begin{center}
\epsscale{0.9}
\rotatebox{90}{
\plotone{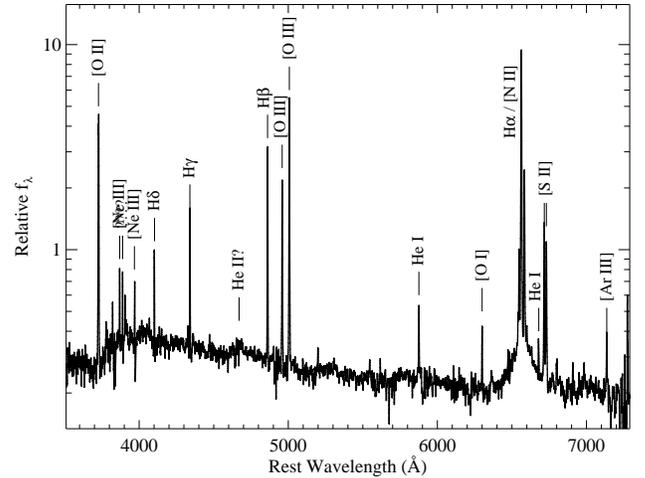}}
\caption[]{Optical spectrum of \LMC902 from 2008 ($t = 1423$~days).
  Narrow emission lines are marked.  The continuum is likely dominated
  by the host galaxy, while the broad H$\alpha$ and potential
  \ion{He}{2} $\lambda 4686$ lines are from \LMC902.
\label{fig:lines}}
\end{center}
\end{figure}

Several narrow galactic emission lines are present in all spectra and
are identified in Figure~\ref{fig:lines} at a redshift of $z =
0.289$.  Throughout this paper we assume a cosmology with $H_{0} =
70$~\kms~Mpc$^{-1}$, $\Omega_{m} = 0.3$, $\Omega_{\Lambda} =
0.7$, which yields a luminosity distance of $d_{L} = 1486$~Mpc for the
SN redshift.

\section{Comparisons to Other Core-Collapse Supernovae}\label{sec:comp}

\subsection{Spectroscopic Comparisons}\label{ss:spec_comp}

The five spectra of \LMC902 are presented in Section~\ref{ss:spec}.
The 2008 and 2009 spectra are dominated by host-galaxy light (see
Figure~\ref{fig:lines}); however, there is a strong, broad component
to the H$\alpha$ line.  We are confident that this broad component is
from the underlying SN and not from another source such as an AGN (for
instance, the broad component is not visible in our 2003 spectrum).

The 2003 spectrum is much bluer than the late-time spectra, indicating
that it contains more SN light than the late-time spectra, which is
not surprising considering that the SN was \about 3~mag brighter in
2003 compared to 2008.  In order to isolate the SN continuum in the
2003 spectrum, we have subtracted a galaxy template spectrum from
the 2003 spectrum.  We chose the galaxy template to match the general
continuum shape of the late-time spectra.  The pre-event photometry
can be used to properly scale the galaxy template.  Similarly, the
photometry at the time of each SN spectrum can be used to scale each
spectrum.  Unfortunately, this procedure produces a galaxy spectrum
that is \about 10\% too bright compared to the late-time spectrum
(assuming that the late-time spectrum is dominated by the host
galaxy).  However, considering the fact that the galaxy template
spectrum does not include emission lines and there may be crowding in
the template image, we do not consider this to be a significant
discrepancy.  The galaxy template spectrum was rescaled to match the
continuum of the 2008 spectrum and subtracted from both the 2003 and
2008 spectra.

\begin{figure}[t]
\begin{center}
\epsscale{0.9}
\rotatebox{90}{
\plotone{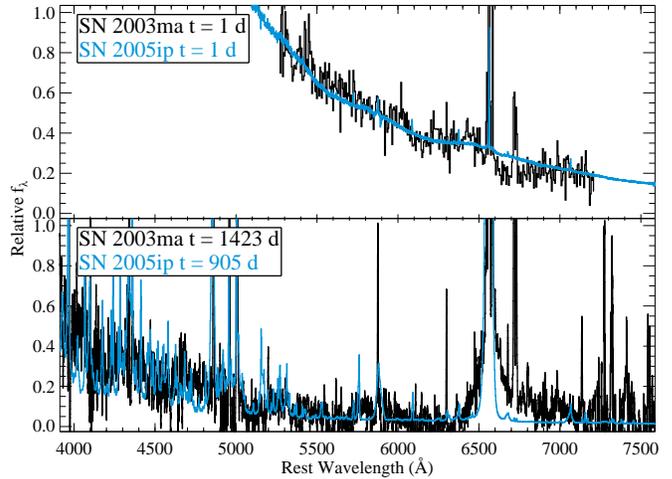}}
\caption[]{Optical spectra of \LMC902 compared with those of
  SN~2005ip.  Phases are given in rest-frame days since maximum light. 
  Top: Galaxy-template subtracted maximum-light spectrum of \LMC902
  (black) and maximum-light spectrum of SN~2005ip (blue).  An
  additional E(B-V)=0.9~mag of internal extinction was applied to
  SN~2005ip to match the continuum shape of \LMC902.
  Bottom: The
  1423~day (4~year; black) and 905~day (2.5~year; blue) spectra of
  \LMC902 and SN~2005ip, respectively.  A template galaxy spectrum has
  been added to the SN~2005ip spectrum to match the continuum shape of
  the \LMC902 spectrum.
\label{fig:spec_comp}}
\end{center}
\end{figure}

We present the dereddened (our spectra of \LMC902 have only been
corrected for foreground extinction based on Galactic dust maps
\citealt{Schlegel98}; see Section~\ref{sec:extinction})
galaxy-subtracted 2003 and 2008 spectra in Figure~\ref{fig:spec_comp}.
We compare these spectra to those of SN~2005ip \citep{Smith09_05ip},
after dereddening the spectra of SN~2005ip by $E(B-V) = 0.047$~mag to
account for Milky Way extinction.  In order to match the continuum
slope of \LMC902, we add an additional $E(B-V) = 0.9$~mag of host
extinction.  \citet{Smith09_05ip} found that SN~2005ip may have had
host extinction of $E(B-V) = 0.3$~mag based on measurements of Na~D
absorption, however since the absorption features were only clearly
detected in the spectrum taken 1 day after discovery, they were unsure
of the true magnitude of extinction.  It seems unlikely that SN~2005ip
suffered from $E(B-V) = 0.9$~mag of reddening without strong Na~D
absorption.  Instead, the difference in continuum shapes may indicate
that \LMC902 was hotter at this epoch than SN~2005ip (see also
Section~\ref{sec:peakT}).

At maximum, both \LMC902 and SN~2005ip have relatively featureless
continua.  With the applied extinction corrections, there are no
significant deviations between the two spectra. The spectrum of
\LMC902 is consistent with that of a blackbody.  There is a low
amplitude, broad feature in the \LMC902 spectrum at \about 6150~\AA
(in the rest frame) that may correspond to H$\alpha$ absorption.  If
this identification is correct, the feature would have an absorption
minimum blueshifted by \about 20,000~\kms.  In
Figure~\ref{fig:spec_comp}, we also compare the 2008 spectrum of
\LMC902 to the $t = 905$~day spectrum of SN~2005ip
\citep{Smith09_05ip}.  The continua of the spectra match, but this may
be the result of an incorrect treatment of the galaxy contamination.
The only feature of significance is the H$\alpha$ line, which is much
broader for \LMC902 compared to SN~2005ip.  The broad H$\alpha$
line is also seen in the 2009 spectrum, but at lower equivalent
width.

We detect narrow \ion{He}{1} $\lambda 5876$ emission in the spectra
from 2004, 2008, and 2009, but do not detect it in the 2003 spectrum
(the wavelength range of the 2005 spectrum does not cover this
feature).  The 2003 spectrum has a relatively low S/N, and the
\ion{He}{1} may be too weak to be detected.  Alternatively, in
Section~\ref{sec:peakT}, we argue that the blackbody temperature in
2003 was $\gta 15,000$~K, which should produce sufficient high-energy
photons to singly ionize He.  In Figure~\ref{fig:lines}, we
tentatively identify relatively broad \ion{He}{2} $\lambda 4686$ in
the 2008 spectrum.  If correct, this would require a significant
amount of ionizing far-UV or X-ray photons from the circumstellar
interaction to ionize He in the post-shock material.  However, as
shown in Figure~\ref{fig:spec_comp}, SN~2005ip has several narrow
lines in this region of the spectrum.  It is possible that the 2008
spectrum of \LMC902 does not have a high enough S/N to see these
individual lines.

In SNe~IIn, the H$\alpha$ line can have up to four distinct
components: broad, intermediate, narrow, and galactic.
\citet{Smith09_05ip} was able to identify the first three regions for
SN~2005ip, while the fourth component was not seen.  Physically, the
broad, intermediate, and narrow components are formed in the
fast-moving SN ejecta, the post-shock circumstellar material, and the
pre-shock circumstellar material, respectively.  The widths of the
broad and narrow components indicate the velocity of the SN ejecta and
the circumstellar wind/ejection, respectively, while the intermediate
component is an indication of the amount of energy transferred from
the shock to the circumstellar material.  The galactic component,
which is the line-of-sight emission from the host galaxy and
physically unrelated to the SN, may have a similar velocity to that of
the narrow component.

\begin{figure}[t]
\begin{center}
\epsscale{1.27}
\plotone{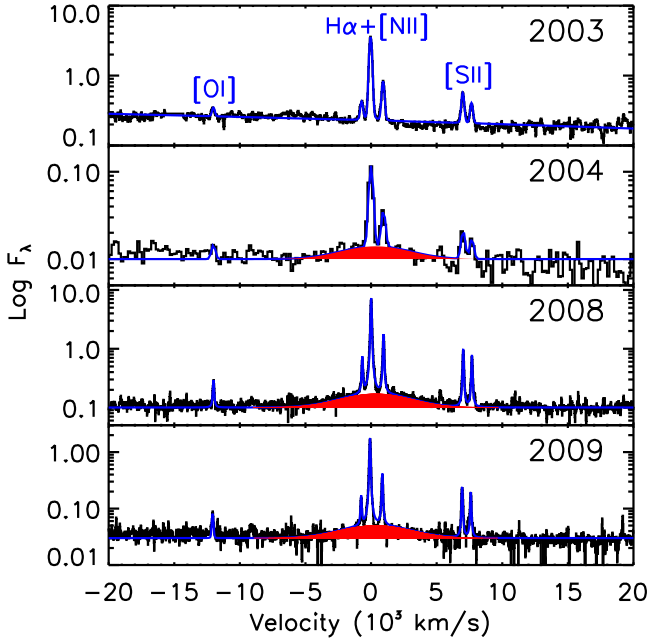}
\caption[]{Narrow [\ion{O}{1}] $\lambda$6300, H$\alpha$, [\ion{N}{2}]
$\lambda\lambda$6548, 6583, and [\ion{S}{2}] $\lambda\lambda$6716,
6731 emission lines fitted with Voigt profiles with a width equal to
the [\ion{N}{2}] $\lambda$6583 line in each spectrum, respectively,
are shown with a blue line.  In 2003 the continuum if fitted with a
quadratic function, and in 2004, 2008, and 2009 the continuum is
fitted with a constant value.  An additional broad Gaussian component
that is needed to fit the H$\alpha$ line in 2004, 2008, and 2009 is
shown in red. 
In 2004 the [\ion{N}{2}] $\lambda$6548 line is not detected above the noise, and is
weaker than expected for a line ratio of 3:1 for  [\ion{N}{2}] $\lambda$6583/$\lambda$6548.
\label{fig:sg_ha}}
\end{center}
\end{figure}

In Figure \ref{fig:sg_ha} we show the narrow
[\ion{O}{1}]~$\lambda$6300, H$\alpha$,
[\ion{N}{2}]~$\lambda\lambda$6548, 6583, and
[\ion{S}{2}]~$\lambda\lambda$6716, 6731 emission lines fitted with a
Voigt profile with its width fixed to that of the
[\ion{N}{2}]~$\lambda$6583 line in each spectrum.  The Voigt profile
produces better fits to the wings of the narrow lines than a single
Gaussian.  The narrow lines are unresolved in the 2003 and 2004
spectra, but the high resolution of the 2008 and 2009 spectra
resolves the narrow lines to have a ${\rm FWHM} = 110$~\kms.  In 2004, 2008, 
and 2009 we fit the broad base beneath the H$\alpha +$[\ion{N}{2}]
narrow-line complex with a Gaussian.  The best fit was for a Gaussian
with ${\rm FWHM} = 5800 \pm 800$~\kms\ and a velocity offset of
$+350$~\kms\ relative to the narrow H$\alpha$ line.  
This velocity offset could be due to asymmetries in the line profile.
The broad
component is not detected in the 2003 spectrum, and we place an upper
limit on the presence of this Gaussian component to be $< 0.2$ times
the flux of the narrow H$\alpha$ line.  The ratio of the broad
component to the narrow component of H$\alpha$ is $0.7 \pm 0.1$, 
$0.4 \pm 0.2$, and $0.5 \pm 0.2$  in 2004, 2008, and 2009, respectively.  
The change in this ratio is not statistically significant, and systematic effects such as
different position angles of the slit and varying seeing can also
produce differences.

\begin{deluxetable*}{lccc}
\tablewidth{0pt}
\tablecaption{Diagnostic Narrow-Line Ratios \label{tab:ratios2}}
\tablehead{
\colhead{Year} & \colhead{[OIII]$\lambda5007$/H$\beta$} & \colhead{[NII]$\lambda6583$/H$\alpha$} & \colhead{[SII]$\lambda\lambda6716,6731$/H$\alpha$}
}
\startdata
2003 &  1.98 $\pm$ 0.07 &  0.18 $\pm$  0.01 &  0.21
 $\pm$  0.02 \\
2004 &  2.19 $\pm$ 0.03 &  0.23 $\pm$  0.04 &  0.21
 $\pm$  0.03 \\
2005 &  2.10 $\pm$ 0.07 & \nodata  & \nodata \\
2008 &  2.19 $\pm$ 0.05 &  0.22 $\pm$  0.02 &  0.18 $\pm$  0.01\\
2009 &  1.71 $\pm$ 0.12 & 0.19 $\pm$ 0.02 & 0.17 $\pm$ 0.01
\enddata
\end{deluxetable*}

We measure the standard diagnostic [\ion{O}{3}]/H$\beta$ and
[\ion{N}{2}]/H$\alpha$, and [\ion{S}{2}]/H$\alpha$ narrow-line ratios
\citep{BPT81, Veilleux87}, which are insensitive to extinction and
atmospheric refraction, for each spectrum from the Voigt profile fits.
The narrow-line ratios, listed in Table \ref{tab:ratios2}, are all
consistent with a star-forming galaxy.  This, in combination with the
lack of variability in the densely sampled \mbox{SuperMACHO} and OGLE
difference imaging data over a baseline of 7 years before the SN,
place strong constraints on the presence of an AGN in the host galaxy.
  
If the broad H$\alpha$ component is powered by CSM interaction, then
we expect there to be some contribution to the narrow H$\alpha$ and
H$\beta$ emission lines.  
However, with no observed velocity offset
between the expected narrow component from the SN and the host galaxy
emission lines, 
it is difficult to tell if the SN contributes to the narrow emission
lines.
Furthermore, the narrow-line ratios
are measured to be constant within the errors from 2003 to 2008, and
thus the lines are most likely dominated by the \ion{H}{2} regions in
the host galaxy.
The luminosity of the H$\alpha$ line in 2003, when the broad component
of H$\alpha$ associated with the SN is not yet observed, is
$L(H\alpha) = 6.1 \times 10^{42}$~ergs~s$^{-1}$, which corresponds to
a substantial star-formation rate of $4.8 M_{\sun}$~year$^{-1}$
characteristic of a starburst galaxy \citep{Kennicutt1998}.



At day 905, SN~2005ip had been on its plateau for over 2 years, and
the broad component of its H$\alpha$ line had faded by a factor of
\about 6 while its narrow+intermediate H$\alpha$ line, which is
assumed to be powered by interaction with the CSM, had increased by a
factor of 3, resulting in a nearly constant total H$\alpha$ luminosity
\citep{Smith09_05ip}.  At this epoch, the FWHM of the intermediate
component and broad component of H$\alpha$ were $740 \pm 250$~\kms\
and $8800 \pm 660$~\kms, respectively \citep{Smith09_05ip}.  In
comparison, at day 279 and 1423, \LMC902 was on its second plateau and
also most likely dominated by circumstellar interaction, however the
width of the broad component of its H$\alpha$ line is at least 5 times
larger than the intermediate component seen in SN~2005ip.  Similarly,
it is significantly broader than the intermediate component of
H$\alpha$ observed in SNe~IIn~1988Z, 2006tf, and 2007rt which had a
characteristic width of 2000~\kms\ \citep{Aretxaga99, Smith08_06tf,
Trundle09} and SN 2006gy which had an intermediate component with
4000~\kms\ \citep{Smith07_06gy}.

The \about 6000~\kms\ component seen in the 2004 and 2008 spectra of
\LMC902 could be associated with SN ejecta that has decelerated over
time.  The width of the broad component of H$\alpha$ was observed to
decrease from \about 15,000~\kms\ on day 34 to 5000~\kms\ after day
500 in SN~1988Z \citep{Aretxaga99} and decrease from \about
10,000~\kms\ on day 102 to $<6000$~\kms\ after day 273 in SN~2007rt
\citep{Trundle09}.  It may be that the day 1 spectrum of \LMC902 was
too early, and the day 279 spectrum was too late to catch the broad
component when it had a velocity width of tens of thousands of \kms.
However, given the long-lived plateau in the light curve,
circumstellar interaction must be the dominant source of energy after
day 90.  Thus, the broad component in \LMC902 is most likely an
exceptionally fast ``intermediate component'' that is the result of
the interaction with the CSM of a highly energetic explosion, even
more energetic than SN 2006gy (see Section
\ref{sec:energydiscussion}), which imparts a much faster blast-wave
speed to the postshock shell. However, electron scattering can also be
a significant source of broadening of the H$\alpha$ line in
interacting SNe, and thus the speed of the material may be slower than
inferred from the width of the line \citep{Dessart2009}.

\subsection{Photometric Comparisons}\label{ss:phot_comp}

The I-band light curve of \LMC902 has two nearly flat plateaus, and
strong emission in the $V$ and $I$ bands for \about 1000~days,
reminiscent of the long-lived emission seen in interacting SNe~IIn.
Although the first plateau in the rest-frame R-band is similar to the
initial plateau seen in SNe~IIP, its comparable plateau in the
rest-frame B band is unlike IIP SNe which begin their decline
immediately after the peak in the B band.  Figure \ref{fig:compall}
shows a comparison of \LMC902 to canonical SN~IIP~1999em
\citep{Leonard02,Elmhamdi03}, along with extreme interacting
SNe~IIn~1988Z \citep{Aretxaga99} and 2005ip \citep{Smith09_05ip}, and
extremely luminous SN~IIn~2006gy \citep{Ofek07, Smith07_06gy},
SN~IIL~2005ap \citep{Quimby07_05ap}, SN~IIL~2008es
\citep{Miller09_08es,Gezari09_08es}, and normal SN~Ia~1988aq
\citep{Riess05}.  All light curves are corrected for Galactic
extinction, and in the case of SN~1999em and SN~2006gy for internal
extinction \citep{Leonard02,Smith07_06gy}. K-corrections are applied
to SNe at distances farther than 100 Mpc.

In SNe IIP, a plateau is observed in the V and R bands as a result of
H recombination in the SN ejecta.  The plateau is not seen at shorter
wavelengths ($U$ and $B$ bands) due to sensitivity to the declining
temperature and line blanketing from metals, nor at longer wavelengths
($I$ band) since the bandpass is on the Rayleigh-Jeans tail of the
emission the whole time \citep{Leonard02}.


The \mbox{SuperMACHO} data constrains the pre-maximum light curve of
\LMC902 exceedingly well (see Figure~\ref{fig:comp88z}). The first
significant detection is 4.5~mag below peak 20~days before maximum,
and we estimate the day of explosion to be between
the last upper limit and the first detection at $t_{0} = {\rm MJD}\
52,971 \pm 1$.  The slow decline of the $I$ band (rest-frame $R$ band)
in the first plateau ($53010 \le {\rm MJD} < 53110$ or
$t-t_{0}=40-140$~days) of $0.0017 \pm 0.0009$~mag~day$^{-1}$ and in
the second plateau ($\rm{MJD} \ge 53110$ or $t-t_{0}=140-1735$~days)
of $0.0011 \pm 0.0001$~mag~day$^{-1}$ is much flatter than the 481 day
decline seen in SN~IIn~2007rt with 0.006~mag~day$^{-1}$
\citep{Trundle09}, and is more like the extreme plateaus observed in
the light curves of SNe~IIn 1988Z and 2005ip (see
Figure~\ref{fig:comp88z}).  The slow decline in these SNe~IIn light
curves are explained by an excess of emission above the photospheric
emission produced by the interaction of the SN ejecta with
circumstellar material.  The spectra of these SNe demonstrated broad
Balmer emission lines, with a narrow component and no P-Cygni
absorption, suggesting that the emission is confined to an outer shell
\citep{Turatto93}.

\begin{figure}[t]
\begin{center}
\epsscale{1.17}
\plotone{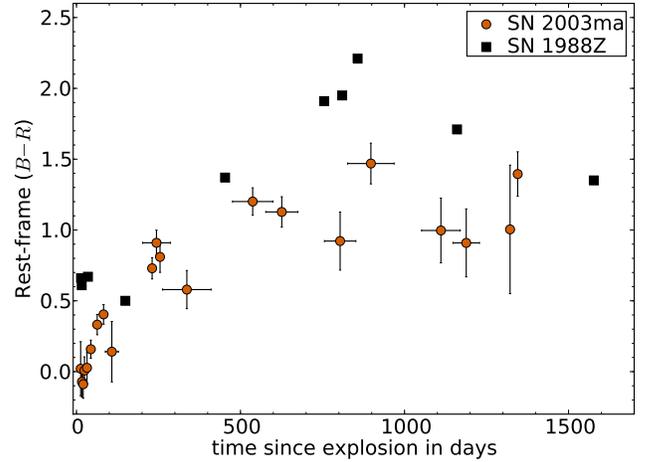}
\caption[]{Rest-frame $B-R$ colors of \LMC902 in comparison to
SN~1988Z in days since the estimated time of explosion $t_{0}$.  The
time of explosion of SN~1988Z is assumed to be 12~days before
discovery. The time axis is in the rest-frame.  Both SNe show an
increase in $B-R$ color until \about 1000~days, after which the $B-R$
color plateaus.
\label{fig:colors}}
\end{center}
\end{figure}

The color evolution of SN~1988Z and \LMC902, shown in Figure
\ref{fig:colors}, are remarkably similar.  The extremely flat plateaus
of this type of SN are attributed to an increase in narrow and
intermediate-width H$\alpha$ flux as well as a forest of other narrow
lines powered by the circumstellar interaction \citep{Smith08_06tf}.
The decreasing H$\alpha$ line velocity width over time in SN~1988Z was
successfully modeled by \citet{Aretxaga99} as the consequence of a
SN remnant shock expanding into a dense medium. The fact that
\LMC902 is bluer than SN~1988Z in particular during the early phases
can be attributed to the higher blackbody temperature.


\section{Physical Parameters of \LMC902}\label{sec:physical}

\subsection{Extinction}
\label{sec:extinction}

\begin{figure}[t]
\begin{center}
\epsscale{0.9}
\rotatebox{90}{
\plotone{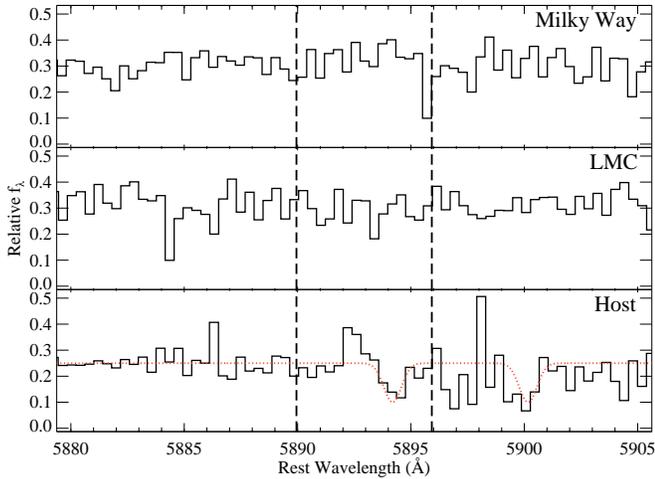}}
\caption[]{Optical spectrum of the 1423~day spectrum of \LMC902 near
  the rest wavelength of the Na~D doublet for the Milky Way (top), LMC
  (middle), and host galaxy (bottom), respectively.  The vertical
  dashed lines indicated the wavelengths of the Na~D lines.  We detect
  no Na~D absorption in the Milky Way or LMC to a equivalent width
  limit of \about 0.4~\AA.  There is the possibility of Na~D
  absorption in the host galaxy with a redshifted velocity of
  200~\kms\ relative to the host galaxy with an equivalent width
  of 0.5~\AA.  This possibility is represented by the red dashed
  lines.
\label{fig:nad}}
\end{center}
\end{figure}

We use the typical galactic extinction law of \citet{Cardelli89}
parameterized by a value of $R_{V} = A_{V}/E(B-V) = 3.1$.  We
obtain\footnote{http://irsa.ipac.caltech.edu/applications/DUST/} a
combined Galactic and LMC reddening of $E(B-V) = 0.348$~mag at the
position of \LMC902 from \citet{Schlegel98}.  As a consistency check,
we examine the spectra of \LMC902 for tracers of interstellar dust.
The Na~D doublet is not detected at or near zero velocity in any of
our spectra.  The 2008 spectrum of \LMC902, which has a resolution of
\about 20~\kms, is ideal for examining the Na~D doublet.  We present
the wavelength regions from that spectrum which correspond to the
rest-frame Na~D doublet for the Milky Way, LMC, and host galaxy of
\LMC902 in Figure~\ref{fig:nad}.  From this spectrum, we can rule out
Na~D absorption with an equivalent width of $\gtrsim 0.4$~\AA\ from
the Milky Way and the LMC.  Using the relationship of
\citet{Barbon90}, we expect $E(B-V) \lesssim 0.1$~mag from the Milky
Way and the LMC, respectively.  However, \citet{Blondin09} found
$E(B-V) > 0.3$~mag for objects with Na~D equivalent widths below about
0.3~\AA.  We therefore believe that our measurements of Na~D are
consistent with the \citet{Schlegel98} extinction estimates.

In Figure~\ref{fig:nad}, we see that there is a possible detection of
Na~D absorption from the host galaxy at a redshifted velocity of $v =
200$~\kms\ relative to the host galaxy with a width of $\lesssim
20$~\kms, the resolution of our spectrum.  However, we do not see a
narrow emission line component from the SN that is offset from the
emission lines of the host galaxy.  There are three possible
explanations for the lines.  First, the lines could be a spurious
detection caused by random noise.  This is plausible given the level
of noise in the spectrum and the slightly odd line profiles.  Second,
there could be an absorbing cloud traveling toward \LMC902.  \LMC902 is at
the same redshift as its host galaxy; therefore, \LMC902 (either in a
disk or the halo) would have its velocity vector pointing mostly
perpendicular to our line of sight. If the cloud is in the disk, then
\LMC902 must be in the halo, and likewise, if \LMC902 is in the disk,
then the cloud must be in the halo.  Third, there is local material
falling inward towards \LMC902.  Either circumstellar material or gas
from an \ion{H}{2} region may be under gravitational collapse, however
the velocity appears to be very high for such a scenario.  Given the
possibilities, we believe the most plausible is that there is no
significant Na~D absorption near the redshift of the host galaxy,
which indicates no additional extinction from the host
galaxy. However, the fact that there is a strong IR echo (see
Section~\ref{sec:dust}) in combination with the type of the host galaxy
being a star-forming galaxy which are generally dust-rich, makes it
very likely that there is significant extinction in the host. The lack
of Na~D absorption might then be due to bleaching of the Na~D doublet
out to a few hundred pc by photoionizing \ion{Na}{1} to
\ion{Na}{2}. Considering all these circumstances we assume as a lower
limit no additional extinction from the host galaxy throughout the
paper. Consequently, the total assumed $E(B-V) = 0.348$~mag is also a
lower limit. Any additional extinction beyond the Milky Way Galaxy and
LMC foreground component increases the required SN luminosity and
hence makes the SN even more extreme in its characteristics.

For each passband, we compute the combined Galactic and LMC extinction
at its effective wavelength $\lambda_{\rm eff}$ with $A(\lambda_{\rm
eff}) = \mean{A(\lambda_{\rm eff})/A_{V}} E(B-V) R_{V}$ using
Equation~1 from \citet{Cardelli89}. Table~\ref{tab:kcorrections_solar}
shows the effective wavelengths and extinction for the various
passbands in the second and third column, respectively.

\subsection{Temperature at Peak}
\label{sec:peakT}

As described in Section~\ref{ss:spec_comp}, we obtained a spectrum of
the event and host galaxy on 2003 December 22, \about 3~days after
peak.  To examine the temperature of this spectrum, we focus on the
dereddened, galaxy-subtracted spectrum (see Section~\ref{ss:spec_comp}
and Figure~\ref{fig:spec_comp}).  We ignore the region of the spectrum
corresponding to 6350 -- 6800~\AA\ in our analysis below to avoid any
H$\alpha$ contamination.  The spectrum shows a blue continuum without
any significant emission or absorption lines. This is similar to the
spectra of SNe~II caught within days of explosion which also show a
relative featureless blue continuum which can be fitted with blackbody
temperatures of \about 10,000 -- 20,000~K \citep[e.g.,][]{Dessart06,
Dessart08, Gezari09_08es}.  The earliest spectrum of SN~2005ip also
demonstrates a blue, featureless continuum on day 1 after discovery
that is fitted with a blackbody with $T_{\rm BB} = 7300$~K; however,
this temperature is most likely a lower-limit due to the uncertainty
in the amount of internal extinction \citep{Smith09_05ip}.

\begin{figure}[t]
\begin{center}
\epsscale{1.15}
\plotone{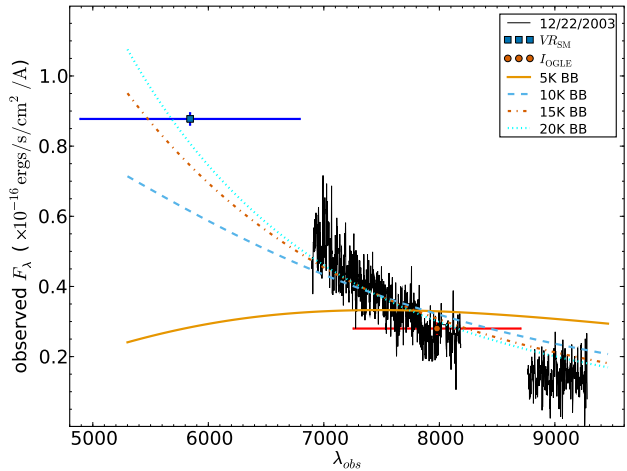}
\caption[]{Spectrum of \LMC902 (black line) from 2003 December 22
  (\about 3~days after peak) with galaxy contamination subtracted as
  discussed in Section~\ref{ss:spec_comp} and dereddened.  The galaxy
  H$\alpha$ line does not fully subtract.  Therefore the spectrum in
  the rest frame wavelength range 6350 -- 6800~\AA\ is not reliable
  and we do not use it in the analysis.  Overplotted are the
  dereddened observed fluxes derived from the \VR$_{\rm SM}$ (blue
  squares) and $I_{\rm OGLE}$ (red diamonds) photometry. The
  horizontal error bars indicate the width of the filters. The solid
  orange line, dashed sky blue line, dot-dashed purple line, and
  dotted cyan line indicate blackbody fits of 5000, 10,000, 15,000,
  and 20,000~K to the spectrum and photometry, respectively.
\label{fig:BBfit}}
\end{center}
\end{figure}

Figure~\ref{fig:BBfit} shows the 2003 galaxy-subtracted spectrum of
\LMC902 (black line), in comparison to black bodies of 5000, 10,000,
15,000, and 20,000~K.  In order to further constrain the blackbody
fit, we use the observed broadband fluxes in \VR$_{\rm SM}$ and
$I_{\rm OGLE}$ (blue square and red diamond).  The horizontal error
bars indicate the width of the filters. Black bodies with $T \lta
10,000$~K do not fit well (see solid orange line and dashed sky blue).
For higher temperatures, a differentiation is not possible since the
observed wavelength range only covers the Rayleigh-Jeans tail of the
blackbody. Therefore we can only set a lower limit of the temperature
at peak of $T_{\rm peak} \gta 15,000$~K.

\subsection{Absolute Magnitudes}
\label{sec:absmag}

\begin{figure*}[t]
\begin{center}
\epsscale{1.17}
\plotone{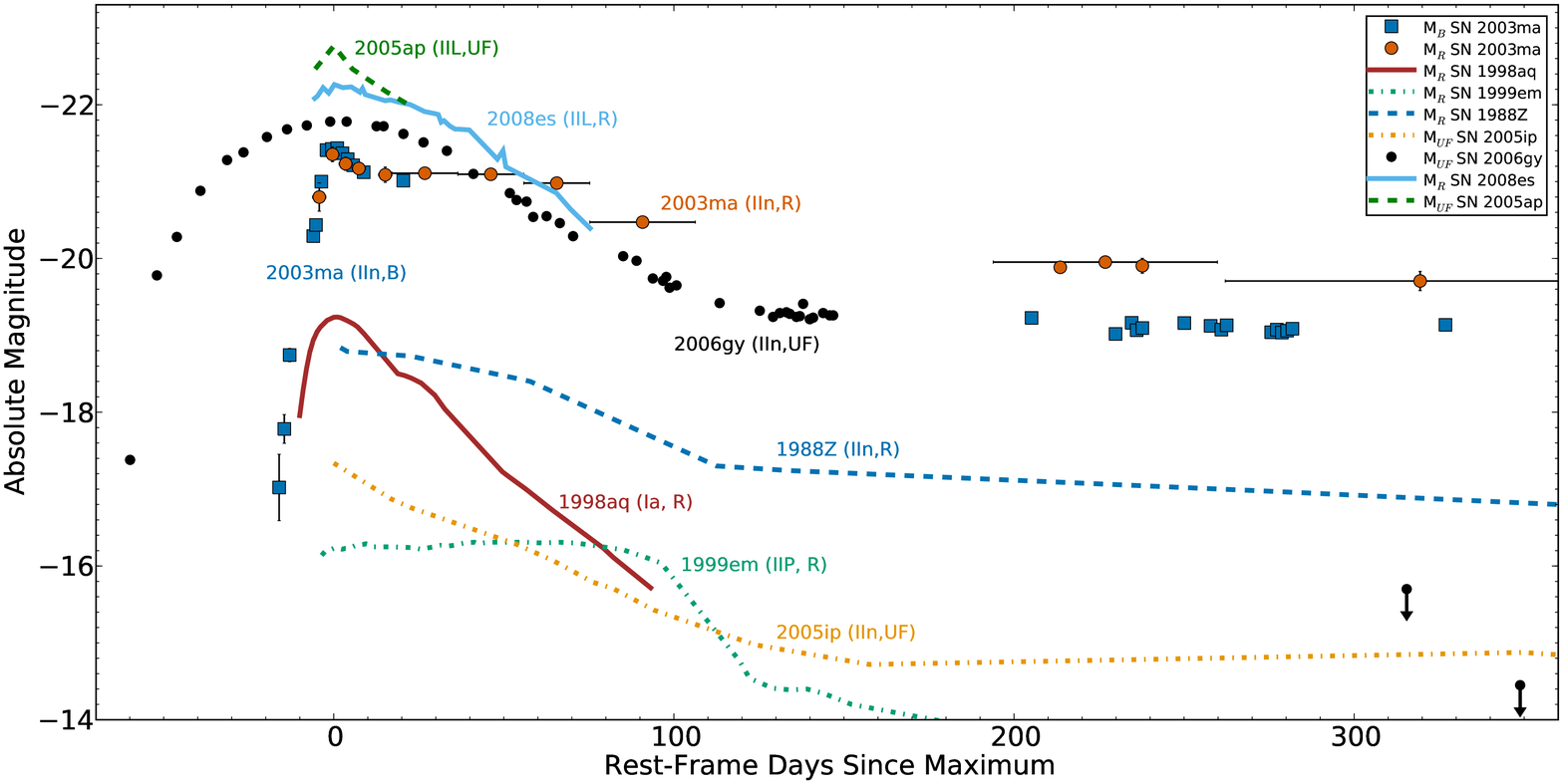}
\caption[]{Light curve of \LMC902 ($DM = 40.86$~mag, $z =
0.289$) in absolute, rest-frame $B$ and $R$ band magnitudes in
comparison to SN~IIn~1988Z \citep[$DM =
34.77$~mag;][]{Aretxaga99}, SN~IIn~2005ip \citep[$DM =
32.36$~mag;][]{Smith09_05ip}, the extremely luminous SN~IIn~2006gy
\citep[$DM = 34.32$~mag;][]{Smith07_06gy, Agnoletto09_06gy},
SN~IIL~2008es \citep[$DM = 40.08$~mag, $z =
0.21$;][]{Miller09_08es, Gezari09_08es}, and SN~IIL~2005ap
\citep[$DM = 40.81$~mag, $z = 0.283$;][]{Quimby07_05ap},
SN~Ia~1998aq \citep[$DM =31.696$,][]{Riess05}, and SN~IIP~1999em
\citep[$DM = 29.569$~mag;][]{Leonard02} in rest-frame days since
maximum.  The times of maxima for SNe~1998aq, 1999em, 2006gy, 2008es,
and 2005ap are well constrained, but for SNe~1988Z and 2005ip there
are no upper limits before the date of discovery, and so we assume
that the date of discovery is the peak date.  The light curves are in
rest-frame $B$, $R$ or $R$-like unfiltered ($UF$)
magnitudes\footnote{The upper limits on the brightness of SN~2006gy
are $R$ band measurements from \citet{Agnoletto09_06gy}}.  We do not
apply any K-correction to the SNe with distances $\lta 100$~Mpc
(SNe~1998aq, 1988Z, 1999em, 2005ip, and 2006gy).  We convert the light
curve of SN~2008es from observed $i$ band to rest-frame $R$ band by
applying a K-correction of $-0.043$~mag. We convert the light curve of
SN~2005ap from unfiltered magnitudes to rest-frame $R$ band by
applying a K-correction of $0.054$~mag, assuming that the observed
magnitudes are in $R$ band.  We apply extinction corrections of 0.036,
0.24, 0.126 and 1.78~mag to the light curves of SNe~1998aq, 1999em,
2006gy and 2005ip, respectively.  The published light curves of
SNe~1988Z, 2008es, and 2005ap are already extinction corrected.  The
observed \VR\ and $I$-band measurements of
\LMC902 are extinction corrected by 0.66 and 1.01~mag, respectively, and
then converted into rest-frame magnitudes in $B$ and $R$ bands with
K-corrections of $-0.63$ and $-0.51$~mag, respectively.  All
K-corrections are calculated assuming a solar spectrum.
\label{fig:compall}}
\end{center}
\end{figure*}

\begin{figure}[t]
\begin{center}
\epsscale{1.17}
\plotone{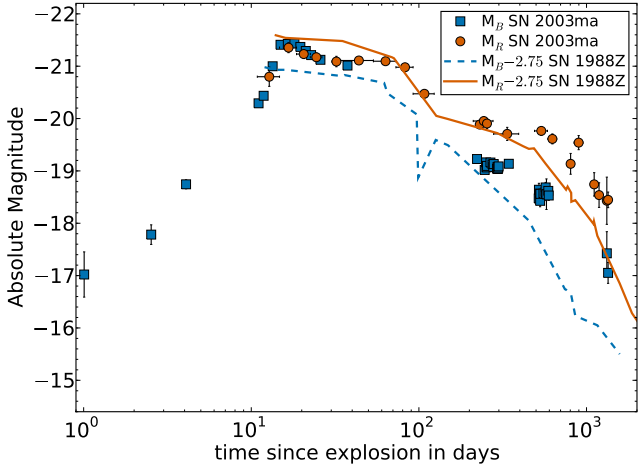}
\caption[]{Light curve of \LMC902 in absolute magnitudes in comparison
to SN~1988Z in days since explosion, on
a log scale. The time axis is in the rest-frame.  Offsets have been
added to the SN~1988Z absolute magnitudes in order to compare the
shape of the light curves. Since the time of explosion is not well
constrained for SN~1988Z, we shift the light curves by 12~days to
match the bend in the $I$ band light curve of \LMC902 seen after
100~days.
\label{fig:comp88z}}
\end{center}
\end{figure}

We calculate the absolute magnitude $M_{X}$ in the rest-frame filter $X$
from an observed magnitude $m_{Y}$ in filter $Y$ as
\begin{equation}
M_{X} = m_{Y} - A_{Y} - DM - K_{XY}
\end{equation}
where $A_{Y}$ is the extinction in filter $Y$ as described in
Section~\ref{sec:extinction}, $DM$ is the distance modulus and
$K_{XY}$ is the K-correction.  For a redshift of $z=0.289$, we
compute $DM = 40.86$~mag. At this redshift, the filters $B$, \VR, and
$I$ in the observers frame are best matched by the rest-frame filters
$U$, $B$, and $R$, respectively.  The K-correction $K_{XY}$ as well as
the bolometric correction described in the next section depend on the
choice of the spectral energy distribution (SED) used to represent the
SN SED.  Due to the limited wavelength range of the observed spectra
in addition to the imperfect background subtraction in particular for
the later time spectra, we cannot use the \LMC902 spectra to directly
calculate the bolometric correction.  Thus, we choose the following
two SED models:
\begin{itemize}
\item {\it \bf Solar model:} The solar spectrum is often
  used to estimate the bolometric correction.  Although it does not
  necessarily represent the true underlying SED, it is commonly used
  in the literature for other SNe, and thus it allows for a direct
  comparison to them.
\item {\it \bf \BB15K88z model:}
  As we show in Section~\ref{sec:peakT}, the peak temperature is
  $T_{\rm peak} \gta 15,000$~K.  Thus we choose a 15,000~K blackbody
  as our SED during the peak (${\rm MJD} < 53010$).  Motivated by the
  fact that the light curves and color evolution of \LMC902 are very
  similar to SN~1988Z after peak, we use the 6 April 1989 spectrum of
  SN~1988Z as the SED for the first plateau ($53010 \le {\rm MJD} <
  53110$), and an average of three SN~1988Z spectra (corresponding to
  dates 18 April 1990, 20 February 1991, and 4 February 1992) for the
  second plateau (${\rm MJD} \ge 53110$).  The particular SN~1988Z
  spectra were chosen for their large wavelength coverage (4500 --
  9000~\AA) and high S/N.
\end{itemize}
While the solar SED is simple and should provide reasonable
measurements, the \BB15K88z model is the more physical model since it
attempts to accurately describe the SN SED.  The fifth column in
Table~\ref{tab:kcorrections_solar} and the third through fifth columns
in Table~\ref{tab:kcorrections_BB15K88z} show the K-corrections for
the two models.  Note that the differences between the K-corrections
derived from the different models are a few tenths of a magnitude.
Figure~\ref{fig:compall}~and~\ref{fig:comp88z} show the absolute
magnitude light curve of \LMC902 (blue squares and red circles)
derived assuming the solar SED for the K-corrections.

%
%

%

\subsection{Luminosity \& Energy}
\label{sec:energy}

We derive the luminosity $L_{X}$ of \LMC902 in a passband $X$ and its
bolometric luminosity $L$ by comparing it to the sun with
\begin{align}
L_{X} &= L_{\sun, X} 10^{-0.4 \left( M_{X} - M_{\sun, X} \right ) } \\
L_{{\rm bol}, X} &= b_{X} L_{X},
\end{align}
where $b_{X}$ is the bolometric correction for band $X$.  The solar
luminosities $L_{\sun, X}$ are calculated with synthetic photometry
off the solar SED \citep{Lejeune97} through Bessell passbands, and the
solar magnitudes $M_{\sun, X}$ are taken from \citet{Colina96}.  They
are shown in the 6th and 7th column in
Table~\ref{tab:kcorrections_solar}, respectively.  We combine $I_{\rm
SM}$ and $I_{\rm OGLE}$ in all rest-frame measurements since they
match well to the same rest-frame passband.  We calculate the
bolometric correction $b_{X}$ for a filter $X$ as the ratio of the
bolometric luminosity of the reference SED with its luminosity in $X$,
where the reference SED is defined by the model used (either solar
model or \BB15K88z model) and for the \BB15K88z model also by its
light-curve phase (see 8th~column in
Table~\ref{tab:kcorrections_solar} and 6th -- 8th column in
Table~\ref{tab:kcorrections_BB15K88z}). Note that the bolometric
luminosity $L_{{\rm bol}, X}$ depends strongly on the bolometric
correction and thus on the filter $X$. Since we have good light curves
in rest frame $B$ and $R$, we can get estimates of the bolometric
luminosity from both light curves and then compare them. In addition,
we add the luminosities in the rest frame $B$ and $R$ together
($L_{B+R}$), and estimate the bolometric luminosity $L_{{\rm bol},
B+R}$ using a bolometric correction $b_{B+R}^{-1} = b^{-1}_{B} +
b^{-1}_{R}$.  Combining both filters decreases the systematic error
since the overall corrections are smaller, as we discuss in
Section~\ref{sec:energydiscussion}.

For both the solar model and the \BB15K88z model we estimate the
amount of energy emitted in the rest frame $B$, $R$, and the combined
$B+R$ passbands by integrating over time (see 10th and 9th column in
Tables~\ref{tab:kcorrections_solar}~and~\ref{tab:kcorrections_BB15K88z},
respectively). The biggest source of systematic error is caused by the
gap in the data in the rest frame $B$ (observed frame \VR) between
$53020 < {\rm MJD} < 53250$, which contains the transition between
plateau 1 and plateau 2 (see Figure~\ref{fig:apparentmags}). A simple
linear interpolation over the gap may introduce a systematic
overprediction of the flux. We estimate that this overestimate is
smaller than 10\%, and does not change our results. Note that the
bolometric correction using the \BB15K88z model is only
pseudo-bolometric, since the SN~1988Z spectra used only have a finite
wavelength range from about 4000 -- 10,000~\AA, not considering any
contribution redward of the $z$ band or blueward of the $B$ band.

\begin{deluxetable*}{cccccccccc}
\tabletypesize{\scriptsize}
\tablecaption{
\label{tab:kcorrections_solar}}
\tablehead{
\colhead{Observed} & \colhead{$\lambda_Y$} & \colhead{$A_Y$} & \colhead{Restframe} & \colhead{$K_{XY}$} & \colhead{$L_{\odot,X}$} & \colhead{$M_{\odot,X}$} & \colhead{$b_X$} & \colhead{$E_{X}$}& \colhead{$E_{{\rm bol},X}$}\\
\colhead{Passband $Y$} & \colhead{(\AA)} & \colhead{(mag)} & \colhead{Passband $X$} & \colhead{(mag)} & \colhead{($10^{32}$~ergs~s$^{-1}$)}  & \colhead{(mag)} & & \colhead{($10^{50}$~ergs)} & \colhead{($10^{50}$~ergs)}  }
\startdata
$B_{\rm SM}$ & 4379 & 1.44 & $U$ & \phantom{+}0.84 & 1.89 & 5.60 & 20.40 \\
$VR_{\rm SM}$ & 5842 & 1.01 & $B$ & $-0.63$ & 4.41 & 5.47 & \phantom{0}8.72 & 4.76 (0.12) & 41.47 (1.06) \\
$I_{\rm SM}$ & 7898 & 0.66 & $R$ & $-0.49$ & \multirow{2}{*}{6.48} & \multirow{2}{*}{4.46} & \multirow{2}{*}{\phantom{0}5.93} & \multirow{2}{*}{4.48 (0.10)} & \multirow{2}{*}{26.57 (0.60)} \\
$I_{\rm OGLE}$ & 7980 & 0.65 & $R$ & $-0.51$ &  \\
$VR+I$ & & & $B$+$R$ &  & & & \phantom{0}3.53 & 9.24 (0.14) & 32.62 (0.49) \\
\enddata
\tablecomments{Table of K-corrections and bolometric corrections for
  the solar model. For each of the passbands $Y$ (first column), the
  effective wavelength $\lambda_Y$, extinction $A_Y$, the passband $X$
  that matches best in the restframe, the K-correction $K_{XY}$, the
  luminosity $L_{\odot,X}$ of the sun in $X$, the absolute magnitude
  $M_{\odot,X}$ of the sun in $X$, and the bolometric correction
  $b_{X}$ are shown in the 2nd-8th column, respectively. The
  K-correction and bolometric correction are calculated using the
  solar SED.  We combine $I_{\rm SM}$ and $I_{\rm OGLE}$ in all restframe
  measurements since they match well to the same restframe
  passband. The 9th column shows the integrated energy $E_{X}$ in
  passband $X$, and the 10th column the integrated bolometric energy
  $E_{{\rm bol},X}$ derived from the passband $X$ photometry.  The
  uncertainties quoted in $E_{X}$ and $E_{{\rm bol},X}$ are the random
  uncertainties and don't contain any of the systematic errors due to
  e.g. the interpolation of the lightcurves, K-correction, etc.}
\end{deluxetable*}

\begin{deluxetable*}{cc|ccc|ccc|cc}
\tabletypesize{\scriptsize}
\tablecaption{
\label{tab:kcorrections_BB15K88z}}
\tablehead{
\colhead{Observed}  & \colhead{Restframe} &  & \colhead{$K_{XY}$} & &  & \colhead{$b_X$} & & \colhead{$E_{X}$}& \colhead{$E_{{\rm bol},X}$}\\
\colhead{Passband $Y$}  & \colhead{Passband $X$} & \colhead{Peak} & \colhead{Plateau 1} & \colhead{Plateau 2} & \colhead{Peak} & \colhead{Plateau 1} & \colhead{Plateau 2} & \colhead{($10^{50}$~ergs)} & \colhead{($10^{50}$~ergs)}}
\startdata
$VR_{\rm SM}$ & $B$ & $-0.44$ & $-0.54$ & $-0.61$ & \phantom{0}7.62 & 26.66 & 27.57 & 5.04 (0.12) & 124.96 (3.40) \\
$I_{\rm SM}$ & $R$ & $-0.57$ & $-0.59$ & $-0.74$ &  \multirow{2}{*}{12.22} &  \multirow{2}{*}{\phantom{0}3.88} &  \multirow{2}{*}{\phantom{0}6.15} &  \multirow{2}{*}{3.69 (0.08)} &  \multirow{2}{*}{\phantom{0}22.28 (0.49)}  \\
$I_{\rm OGLE}$ & $R$ & $-0.57$ & $-0.62$ & $-0.80$ &  &  &  \\
$VR+I$ &  $B$+$R$ &  &  &  & \phantom{0}4.69 & \phantom{0}3.38 & \phantom{0}5.03 & 8.67 (0.13) & \phantom{0}40.15 (0.63) \\
\enddata
\tablecomments{Table of K-corrections and bolometric corrections for
  the \BB15K88z model. For each of the passbands $Y$ (first column),
  the passband $X$ that matches best in the restframe (second column),
  the K-correction $K_{XY}$, and the bolometric correction $b_{X}$ are
  shown. For both the K-correction and bolometric correction, we use 3
  different SED's for the 3 different light curve phases peak,
  plateau~1, and plateau~2. For the peak, we use a 15,000 K blackbody
  SED. Since the lightcurve and color evolution of the first and
  second plateau is similar to that of SN 1988Z, we use a spectrum of
  SN 1988Z from 04/06/1989 as SED for plateau~1, and an average of SN
  1988Z spectra from 04/18/1990, 02/20/91, and 02/04/92 for plateau~2.
  The K-correction for the peak, plateau~1 and plateau~2 is shown in
  3rd, 4th, and 5th column, respectively. The bolometric correction
  for the peak, plateau~1 and plateau~2 is shown in the 6th, 7th, and
  8th column, respectively.  We combine $I_{\rm SM}$ and $I_{\rm OGLE}$ in all
  restframe measurements since they match well to the same restframe
  passband.  The 9th column shows the integrated energy $E_{X}$ in
  passband $X$, and the 10th column the integrated bolometric energy
  $E_{{\rm bol},X}$ derived from the passband $X$ photometry. The
  uncertainties quoted in $E_{X}$ and $E_{{\rm bol},X}$ are the random
  uncertainties and don't contain any of the systematic errors due to
  e.g. the interpolation of the lightcurves, K-correction, etc.}
\end{deluxetable*}

\subsection{Energy Emitted}
\label{sec:energydiscussion}

\begin{figure*}[t]
\begin{center}
\epsscale{1.15}
\plotone{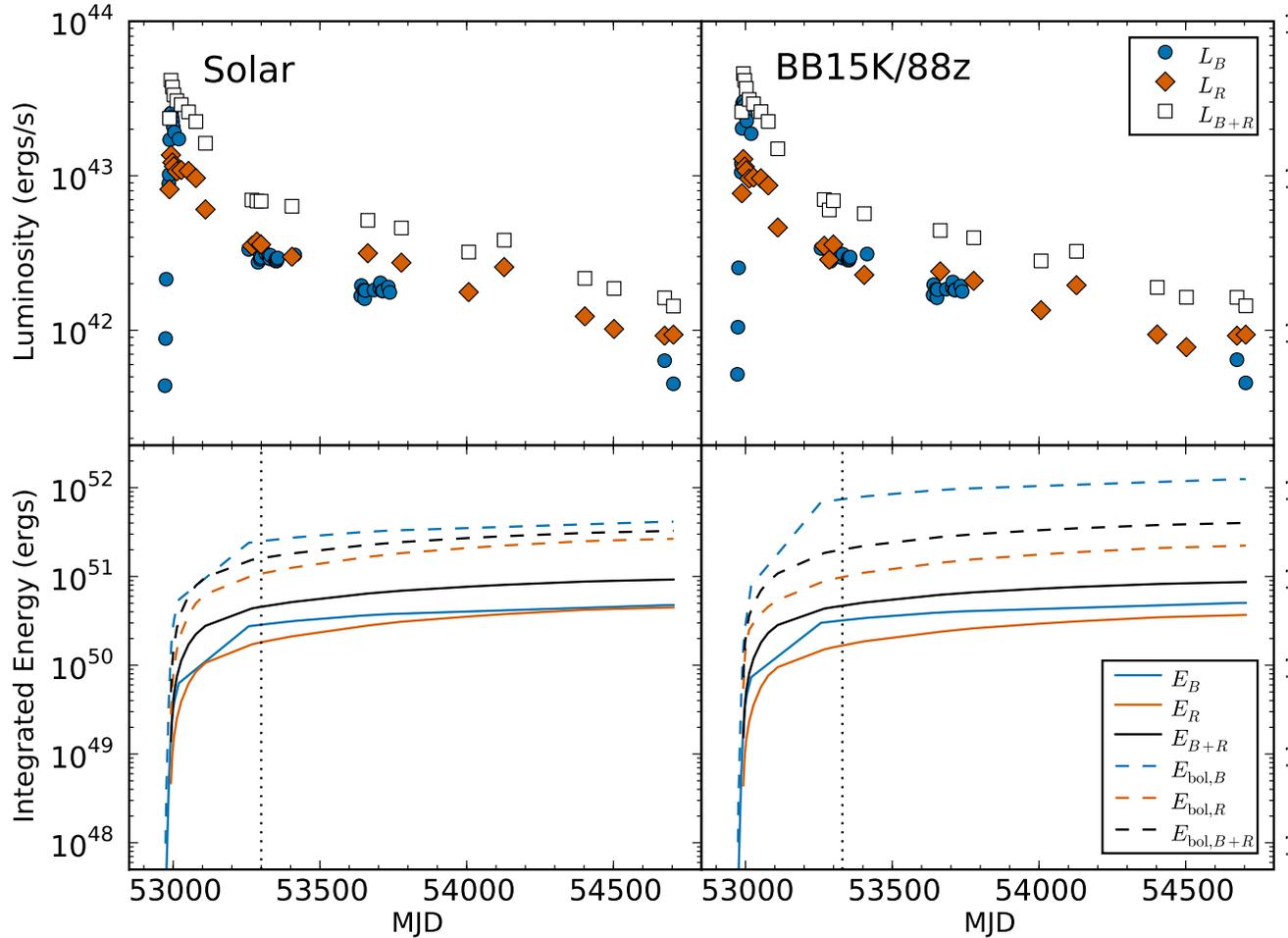}
\caption[]{Top left: The luminosities in the rest frame $B$, $R$ and
combined $B+R$ passbands using the solar SED for the determination of
the K-correction (solar model). Top Right: Same luminosities, but
using a 15,000~K blackbody SED for the K-correction during peak, and
SEDs from SN~1988Z for the K-corrections during the plateau phase
(\BB15K88z model). Bottom left and right: The integrated emitted
energy in the $B$, $R$ and combined $B+R$ passbands using the solar
and \BB15K88z model, respectively, in solid lines. The dashed lines
show the integrated bolometric energy for the two models. See
Section~\ref{sec:energydiscussion} for a discussion.
\label{fig:energy}}
\end{center}
\end{figure*}

The upper left panel of Figure~\ref{fig:energy} shows the luminosities
in the rest frame $B$, $R$ and combined $B+R$ passbands using the
solar SED for the determination of the K-correction. In the upper
right panel, the same luminosities are shown, but using a 15,000 K
blackbody SED for the K-correction during peak, and SEDs from
SN~1988Z for the K-corrections during the plateau phase. The
K-corrections for the two models are listed in
Table~\ref{tab:kcorrections_solar}~and~\ref{tab:kcorrections_BB15K88z}.
Note that the differences in K-corrections and thus luminosity are not
bigger than 20\%.

We estimate the total emitted energy in the different passbands by
integrating over time (see solid lines in lower panels of
Figure~\ref{fig:energy}, and 10th and 9th column in
Tables~\ref{tab:kcorrections_solar}~and~\ref{tab:kcorrections_BB15K88z},
respectively). For the solar model, the emitted energy in rest-frame
$R$ band is $4.5 \times 10^{50}$~ergs, with a similar amount emitted
in the rest-frame $B$ band of $4.8 \times 10^{50}$~ergs.  This is more
than ten times the energy emitted by SN~1988Z in the B band over
roughly the same period of time (\about 4.3~year) of \about $3 \times
10^{49}$~ergs \citep{Aretxaga99}. By adding the luminosities in the
rest frame $B$ and $R$ together, and integrating over the light curve,
we calculate the energy emitted in the combined rest frame $B$ and $R$
to be $E_{B+R} = 9.2 \times 10^{50}$~ergs (black line in lower left
panel of Figure~\ref{fig:energy}). When we use the \BB15K88z model for
the K-correction, we get energies in $B$, $R$, and $B+R$ of $5.0
\times 10^{50}$, $3.7 \times 10^{50}$, and $8.7 \times 10^{50}$~ergs,
respectively. These energies differ by only $\lta 15\%$ from the
energies based on the solar model, and we thus estimate that our
systematic errors are $\lta 15\%$.

The bolometric luminosities and hence bolometric energies have a
strong dependency on the SED used to calculate the bolometric
corrections. The dashed lines in the lower panels of
Figure~\ref{fig:energy} show the bolometric energies derived from the
different passbands using the solar and \BB15K88z model on the left
and right, respectively. The bolometric energies based on the solar
model agree reasonably well and range between $2.7 \times
10^{51}$ and $4.1 \times 10^{51}$~ergs. However, the bolometric
energies derived using the \BB15K88z model significantly differ,
e.g., a factor of five between rest frame $R$ ($2.2 \times
10^{51}$~ergs) and $B$ ($12.5 \times 10^{51}$~ergs), even though this
model is physically more motivated.

This large difference between the rest frame $B$ and $R$ derived
bolometric energy is mainly driven by the bolometric correction based
on the SN~1988Z spectra for the two plateau phases. The late-time
spectra of SN~1988Z-like SN are dominated by the strong H$\alpha$
emission line on the one hand, and the blue pseudo-continuum on the
other hand.  At late phases, this blue continuum is detected in most
interacting SNe, e.g., SN~2006jc \citep{Foley07}, and has been shown
by \citet{Smith09_05ip} to be fluorescence of a forest of emission
lines that can only be resolved if the lines are narrow enough, e.g.,
SN~2005ip. The main difference between the spectral shape of these
late-time spectra of SNe~IIn is the relative strength of the H$\alpha$
emission line (rest frame $R$) compared to the blue pseudo-continuum
(rest frame $B$ and bluer). The rest frame $B-R$ color of SN~1988Z is
significantly redder than the one of \LMC902, indicating that for
\LMC902 the relative strength of the H$\alpha$ line compared to the UV
pseudo-continuum is weaker than for SN~1988Z.  Thus using the spectra
of SN~1988Z leads to a bolometric over-correction in the rest frame
$B$ band and an under-correction in the rest frame $R$ band, which
explains the large discrepancy between the two estimated bolometric
luminosities.  By combining rest frame $B$ and $R$, this effect gets
largely compensated, and the bolometric energy of $4.0 \times
10^{51}$~ergs is comparable to the bolometric energy using the solar
model of $3.2 \times 10^{51}$~ergs. We thus estimate the bolometric
radiated energy of \LMC902 to be $(3.6 \pm 1) \times 10^{51}$~ergs,
with the error dominated by the systematic error in the bolometric
correction. This is significantly more than the total radiated energy
of 2.3 -- 2.5~$\times 10^{51}$~ergs by SN~2006gy in \about 0.6~years
\citep{Smith09_06gy}, and $1.1 \times 10^{51}$~ergs for SN~2008es in
\about 0.4~years \citep{Miller09_08es}, and ten times the energy
radiated by SN~1988Z over 8.5 years of $3.2 \times 10^{50}$~ergs
\citep{Aretxaga99}.  Analogous to SN~1988Z, the true bolometric
luminosity of \LMC902 is probably a factor of 10 higher if you include
emission from the radio to X-rays that is produced from the
interaction of the SN shock with the CSM, resulting in a total
radiated energy budget of at least 10$^{52}$~ergs.

Even though \LMC902 is not quite as bright at peak as the most
luminous SNe like SNe~2005ap, 2006gy, and 2008es, its total emitted
energy is larger since half of the radiated energy is emitted after
350~days due to its very long decline (see the dotted lines in
Figure~\ref{fig:energy} for $E_{{\rm bol}, B+R}$), at times when the
other extremely luminous SNe do not produce any significant radiation.
Such a large amount of radiated energy can be achieved if two
things coincide: (1) the initial explosion has a large total (combined
radiative and kinetic) energy and (2) most of the kinetic energy is
converted into radiation by interaction with a CSM.

\subsection{Peak Luminosity}

The peak luminosity of a SN~II from thermal energy deposited into
the ejecta following shock break out can be estimated to be 
\begin{align}
L_{0} &= \frac{\beta c}{2\kappa}\frac{{E}_{\rm SN}R_{0}}{M} \\
&= 2.5 \times 10^{43}
\left( \frac{E}{10^{51} \rm{ ergs}}\right) 
\left( \frac{R}{10^{14} \rm{ cm}}\right) 
\left( \frac{M_{\sun}}{M}\right) {\rm ~ergs~s}^{-1},
\end{align}
where $E_{\rm SN}$ is the energy of
the SN explosion, $M$ is the mass of the ejecta, $R_{0}$ is the
initial radius of the star, $\beta$ is a numerical constant related to
the radiative diffusion time, and $\kappa$ is the opacity
\citep{Arnett1996}.  Thus, the most luminous core-collapse SNe will
result from progenitors with large initial radii and low densities.
Given the measured total energy emitted in the optical of ${E}_{\rm
SN} = 4 \times 10^{51}$~ergs, and $L_{0} = 5 \times
10^{43}$~ergs~s$^{-1}$, this implies $R_{0} = 2 \times 10^{14}$
$[M/(10 M_{\sun})]$~cm, a value approaching the radii of the largest
known RSGs \citep{Levesque2005}.  If the progenitor were an Eta
Carinae-type massive star with $M \approx 100 M_{\sun}$, this would
require a radius ten times larger, which is difficult to reconcile
with stellar radii predicted by stellar evolutionary models.

If the peak of the light curve is powered solely by radioactive decay,
we expect
\begin{equation}
M_{\rm Ni}/M_{\sun} = \frac{L}{1.42 \times 10^{43}{\rm ~ergs~s}^{-1}} e^{(t/111 {\rm
days})}.
\end{equation}
For \LMC902, which had $L_{\rm peak} = 5 \times 10^{43}$~ergs~s$^{-1}$
at $t = 8.5$~days after explosion in the SN rest-frame, the initial
$^{56}$Ni mass is 4~$M_{\sun}$, which is one to two orders of
magnitude larger than typically produced in SNe~IIP \citep{Hamuy03}.
Although this amount of $^{56}$Ni could be produced in a
pair-instability explosion \citep{Scannapieco05}, the slow rise to
maximum and constant slope of decay after maximum in such models are
incompatible with the light curve of \LMC902.

\subsection{Kinetic Energy}

Measuring the kinetic energy from a SN typically requires an ejecta
mass and a velocity for the ejecta.  For \LMC902 we were unable to
measure the velocity of the ejecta from any of our spectra.  However,
we can estimate the energy converted from shock energy to kinetic
energy by examining the velocity of the post-shock material.  If we
assume that the pre-shock circumstellar material has a coherent
velocity, $v_{\rm w}$ and a mass, $m$, it has a kinetic energy,
\begin{equation}
  E_{0} = \frac{1}{2} m v_{\rm w}^{2}.
\end{equation}
If we assume that the post-shock material has a one-dimensional RMS
velocity, $v_{\rm ps}$, then the post-shock kinetic energy of the
circumstellar material is
\begin{equation}
  E_{\rm ps} = \frac{3}{2} m v_{\rm ps}^{2},
\end{equation}
and the shock deposited an energy,
\begin{equation}\label{e:de}
  \Delta E = \frac{1}{2} m (3 v_{\rm ps}^{2} - v_{\rm w}^{2}).
\end{equation}
If we assume that the pre-shock circumstellar material has a constant
density, $\rho$, and that the shock traveled at a velocity of $v_{\rm
s}$ with no appreciable deceleration, then
\begin{equation}\label{e:mass}
  m = \frac{4}{3} \pi \rho (v_{\rm s} t)^{3}
\end{equation}
is the mass of circumstellar material swept up by the shock in a time
$t$ after explosion.

Combining Equations~\ref{e:de} and \ref{e:mass} and assuming that
$v_{\rm ps} \gg v_{\rm w}$, we find
\begin{equation}
  \Delta E = 2 \pi \rho v_{\rm ps}^{2} v_{\rm s}^{3} t^{3}.
\end{equation}
If we have a constant conversion of kinetic energy to luminosity, then
$L = \eta \Delta E / t$, where $\eta$ is the efficiency factor.  The
density can then be written as
\begin{align}
  \rho &= 5.6 \times 10^{-20} \frac{1}{\eta} \frac{L}{10^{42}
    {\rm ~ergs~s}^{-1}} \left ( \frac{v_{\rm ps}}{6000
    {\rm ~km~s}^{-1}} \right )^{-2} \notag \\
         &\times \left ( \frac{v_{\rm s}}{20,000 {\rm ~km~s}^{-1}}
    \right )^{-3} \left ( \frac{t}{\rm year} \right )^{-2} \rm{
    ~g~cm}^{-3}. \label{e:density}
\end{align}
>From Equations~\ref{e:mass} and \ref{e:density}, we have the total
swept-up mass,
\begin{equation}
  m = 0.03 \frac{1}{\eta} \frac{L}{10^{42}
    {\rm ~ergs~s}^{-1}} \left ( \frac{v_{\rm ps}}{6000
    {\rm ~km~s}^{-1}} \right )^{-2} 
     \left ( \frac{t}{\rm year} \right ) M_{\sun},
\end{equation}
which is independent of any assumed ejecta velocity.  At the time of
the last spectrum, when $L = 5 \times 10^{42}$~ergs~s$^{-1}$, $v_{\rm
ps} = 6000$~\kms, and $t = 3.9$~years, \LMC902 had swept up
0.6~$M_{\sun}$ of circumstellar material and had at least $6.4 \times
10^{50}$~ergs of kinetic energy if $\eta = 1$.  These values are lower
limits since if $\eta < 1$, the swept-up mass and kinetic energy
increases.

%
%

\subsection{Dust}
\label{sec:dust}

\begin{figure}[h]
\begin{center}
\epsscale{1.15}
\plotone{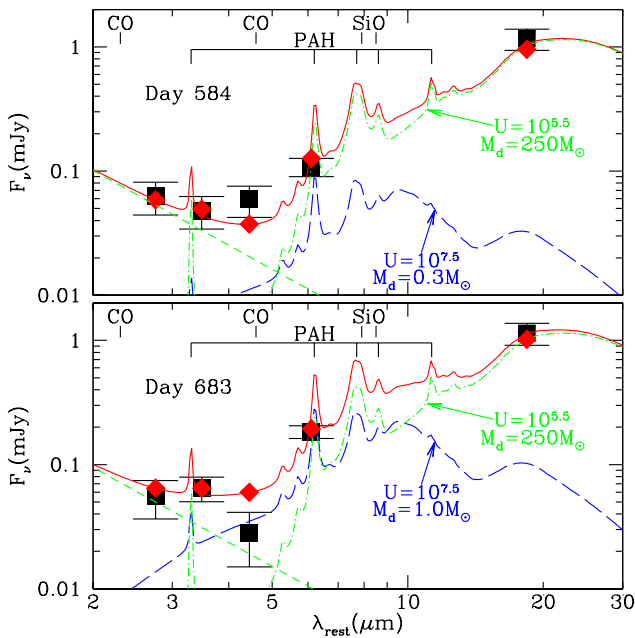}
\caption[]{Spitzer IRAC (3.6, 4.5, 6.0, and 8.0~\micron) and MIPS
(24~\micron) photometry versus rest wavelength, for 584 and 683~days
after maximum.  The solid curves show models which include hot dust,
warm dust (see text), and a stellar component with a Rayleigh-Jeans
spectrum.  Filled diamonds are the model convolved with the IRAC and
MIPs response functions. The photometry is consistent with the
observed emission consisting of steady emission from warm dust in the
galaxy, plus a time-variable IR echo produced by hot dust.
\label{fig:spitzerir}}
\end{center}
\end{figure}

The {\it Spitzer} photometry in IRAC bands 1 -- 4 and MIPS band 1 (see
Table~\ref{tab:spitzer}) is shown in Figure~\ref{fig:spitzerir} with
respect to the rest wavelengths (2.8, 3.5, 4.5, 6.2, and 18.5~\micron,
respectively) corresponding to the observed wavelengths of 3.6, 4.5,
5.8, 8.0, and 24~\micron, respectively. There is a significant
difference between the first epoch 
(2005 July 15 -- 26; \about
580~days after maximum) and the second epoch (2005 October 26 --
November 2; \about 675~days after maximum) at restframe 3.5, 4.5, and
6.2~\micron: 
the fluxes at 3.5 and 6.2 \micron\ increase by 4$\sigma$ and
10$\sigma$, respectively, whereas the 4.5~\micron\ flux decreases by
4$\sigma$.  Note that we use the relative flux error for the
comparison of fluxes at different epochs, but the same filter (4th
column in Table~\ref{tab:spitzer}), as we have discussed in
Section~\ref{sec:IRredux}. For all comparisons of fluxes from
different bands, we use the total errors (6th column in
Table~\ref{tab:spitzer}), which includes the wavelength dependent
error introduced by the crowdedness of the field.

Such a significant change in flux over such a short time span makes it
highly likely that at least some of the measured infrared flux is
caused by the SN. Note, however, that the direction of these
differences across the bands is not in the same direction in
neighboring bands, and therefore \LMC902 would not have been
considered a variable candidate according to the criteria used by
\citet{Vijh09} for the SAGE-LMC variables list. The observed
24~\micron\ flux is essentially unchanged and is either an IR echo as
observed for other SNe (e.g., SN~2002hh, \citealt{Barlow05, Meikle06};
SN~2004et, \citealt{Kotak09_04et}) or from other sources in the host
galaxy.

The changes in flux cannot be explained by a simple change in
temperature and/or flux of a blackbody, but rather requires emission
bands of molecules or dust.  The first detection of carbon monoxide
(CO) in a SN in the fundamental (4.6~\micron) and first overtone
(2.3~\micron) bands was in the spectra of SN~1987A \citep{Catchpole87,
McGregor87, Oliva87, Spyromilio88}. In addition to CO, silicon
monoxide (SiO) emission bands were detected in SN~1987A at 7.9 and 8.5
\micron\ \citep{Roche91, Meikle93, Liu94}. Subsequently, MIR CO lines
were detected in SNe~1995ad \citep{Spyromilio96}, 1998S
\citep{Gerardy00, Fassia01}, 1998dl \citep{Spyromilio01}, 1999em
\citep{Spyromilio01}, 2000ew \citep{Gerardy02b} and 2004et
\citep{Kotak09_04et}, which also showed SiO lines.  The wavelengths of
the CO and SiO emission bands are indicated in
Figure~\ref{fig:spitzerir}.  The match between the CO and SiO lines
with the observed flux changes is poor, except for possible
contribution of the CO fundamental band at 4.6~\micron.  However, we
note that for both SNe~1987A and 2004et the CO and SiO bands appeared
\about 100~days after maximum and disappeared by \about 500~days,
significantly earlier than the increase seen in the second epoch for
\LMC902.

Another possible explanation is that the strong and variable emission
detected in the 8~\micron\ IRAC band is from the 6.2~\micron\ emission
feature of polycyclic aromatic hydrocarbon (PAH) nanoparticles.  Such
emission is routinely detected in the Milky Way and other galaxies
\citep[e.g.,][]{Genzel00, Vogler05, Irwin06, Draine07,
Smith07_MIR_PAH}.

Here we present a dust model that could explain the observed IR
emission.  First, we assume that the bulk of the emission observed at
3.6~\micron\ (rest frame 2.3~\micron) is stellar continuum and we
extrapolate this to longer wavelengths assuming a Rayleigh-Jeans
spectrum (short-dashed line in Figure \ref{fig:spitzerir}).  To this
we add emission from a mixture of PAHs, graphite, and silicate dust,
from the dust models of \citet{Draine07}.  For \LMC902 we assume the
radiation field heating the dust to have the spectrum of a 
15,000~K blackbody, and have solved for the temperature distribution
functions for different grain sizes and compositions, and different
radiation intensities.  The emission model has two components: steady
emission from the ``warm'' dust in the galaxy, plus a time-varying IR
echo from ``hot'' dust heated by radiation from the SN.

To reproduce the 24~\micron\ photometry, the warm dust must be hot
enough for the silicate 18~\micron\ feature to emit strongly.  This is
accomplished by a radiation field with $U \approx 10^{5.5}$ ($U$ is
the dust heating rate relative to heating by the starlight background
in the solar neighborhood).  This dust component is assumed to have
the same ionization fraction as a function of grain size as assumed by
DL07 for the Galaxy, and which has been found to work well for the
IR-submm emission from, e.g., the galaxies in the SINGS sample
\citep{Draine07b}.  The PAH abundance in the warm dust component must
be low, $q_{\rm PAH} \approx 0.5\%$, in order to not overproduce the
observed flux in IRAC bands 2 and 4.  This component has a dust mass
of \about $250 M_{\sun}$; for a normal dust-to-gas ratio, this
corresponds to a hydrogen mass of \about $2.5 \times 10^{4}
M_{\sun}$.  This is a plausible amount of gas for hot
photodissociation regions in a starburst galaxy.  The spectrum of this
emission is shown in Figure~\ref{fig:spitzerir}.  The total IR
luminosity of this dust component is \about $1.1 \times 10^{10}
L_\sun$.  There is presumably additional emission from a much larger
mass of cool dust in the galaxy, but we lack longer wavelength
observations to constrain this component.

The dust producing the IR echo is assumed to be very strongly heated,
with $U \approx 10^{7.5}$.  The IR echo dust mass is \about $0.3
M_{\sun}$ on day 584, and \about $1 M_{\sun}$ on day 683.  The total IR
luminosity of this component on day 683 is \about $4 \times 10^{9}
L_\sun$, requiring that the IR echo dust be absorbing a substantial
fraction of the SN light.  This dust is assumed to have a PAH
abundance $q_{\rm PAH} = 4.7\%$, similar to the PAH abundance in many
solar-metallicity star-forming galaxies.  The PAHs are assumed to be
ionized by the radiation from the SN; the PAH neutral fraction in this
component is assumed to be negligible (the absence of neutral PAHs
lowers the ratio of 3.3~\micron\ emission relative to 7.7 and
8.6~\micron\ emission, which is needed in order to not overproduce the
emission into IRAC band 2).  The intense radiation heats the large
grains to $T \approx 300$~K; the contribution of the broad 9.7~\micron\
silicate feature is apparent in this component of the model spectrum.
Even in this intense radiation field, the emission in the 3.3 and
6.2~\micron\ PAH features continues to be due primarily to
single-photon heating.

This two component model, with the mass of the hot dust increasing by
a factor \about 3 between days 584 and 683, reproduces most of the IR
photometry.  There is admittedly a discrepancy with the observed flux
density in IRAC band 3 ($\lambda_{\rm rest} = 4.5$~\micron): the
observed flux decreases by a factor \about 3 between days 584 and 683,
and our model does not reproduce this.  Perhaps CO 1-0 4.6~\micron\
emission (not included in our model) contributes to the high flux
measured on day 584.

The dust heating parameter $U = 10^{7.5}$ corresponds to an energy
density $u(h\nu < 13.6 {\rm eV}) = 4 \times 10^{-5}~{\rm
erg\,cm}^{-3}$.  The distance of the dust from the SN is not known,
but it is reasonable to assume a distance \about 600 light-days
(0.50~pc).  For this distance, the SN luminosity required to produce
$U = 10^{7.5}$ is \about $1.4 \times 10^{10} L_{\sun}$ -- consistent
with the photometry in Figure~\ref{fig:energy}.

A sphere of radius 0.5~pc with density $n_{\rm H} = 8 \times
10^{3}$~cm$^{-3}$ contains $M_d \approx 1 M_{\sun}$ of dust (assuming
$M_{d}/M_{\rm H} \approx 0.01$).  Densities $n_{\rm H} \gtrsim
10^{4}$~cm$^{-3}$ would be expected for cool gas in the high-pressure
environment of a starburst galaxy.  The IR echo requires \about $1
M_{\sun}$ of dust at a distance \about 0.5~pc from the SN to produce
the IR echo.  This dust could be located in an interstellar cloud, or
possibly in a swept-up shell of interstellar matter surrounding a
stellar-wind bubble.

To summarize: (1) PAH emission features can explain the increase in
the observed 4.5 and 8~\micron\ flux densities; (2) the required dust
mass of $M_d \approx 1 M_{\sun}$ should be \about 0.5~pc from the SN,
which is reasonable; (3) the intensity of the SN radiation at this
distance is consistent with what is required to heat the dust to
reproduce the observed emission.  We conclude that the observed IR
photometry of LMC~902 is consistent with an IR echo produced by normal
interstellar dust at a distance \about 0.5~pc from the SN. 
We note, however, that the data does not sufficiently constrain the
model to make it unique, and other explanations are possible.


\section{Conclusions}\label{sec:conclusions}

This paper presents photometry and spectroscopy of \LMC902, an
exceptional SN~IIn at a redshift of $z=0.289$ behind the Large
Magellanic Cloud.

\LMC902 has an $I$-band light curve with two long plateaus, and
strong emission in the $V$ and $I$ bands for \about 1000~days,
reminiscent of the long-lived emission seen in interacting SNe~IIn
like the prototype of the class, SN~1988Z.  However, in contrast to
all other SNe~IIn with long-lived plateaus, \LMC902 also has an
exceptionally bright peak luminosity of M$_{R} = -21.5$~mag, placing
it in the class of extremely luminous core-collapse SNe, such as
SNe~2005ap, 2006gy, 2008es, and 2008fz.  The combination of high peak
luminosity and long-lasting plateau makes \LMC902 more luminous than
any other SN at 80~days past maximum brightness.  Consequently, half
of the radiated energy is emitted after 350~days past maximum light,
while other extremely luminous SNe produce no significant radiation at
such late times.

The derived bolometric luminosity, and hence bolometric energy, has a
strong dependency on the SED used to calculate the bolometric
corrections.  In order to estimate the systematic bias of the assumed
SED, we use different SEDs to calculate the bolometric corrections.
Integrating the bolometric light curve, we find that a total
bolometric output of \about $4 \times 10^{51}$~ergs was emitted over
4.7~years -- more than twice the total radiated energy of SNe~2006gy or
2008es, and ten times the energy radiated by SN~1988Z, making \LMC902
the most energetic SN ever observed with respect to its radiative
output.  This value alone is within an order of magnitude of the
maximum energy that a core-collapse explosion can produce outside of
the neutrino channel.

The nearly flat late-time light curve over the long time-span of
several years in combination with the persistent single-peaked
intermediate-width H$\alpha$ emission line are best explained by the
conversion of the kinetic energy of the explosion into radiation by
strong interaction of the SN with a dense circumstellar material.
\LMC902 is distinct from proto-typical SNe~IIn in that the
intermediate width line is unusually broad, and it does not have a
broad component. If \LMC902 has narrow emission lines cannot be
answered conclusively since it is difficult to separate the narrow
emission lines from the SN, if they exist, from the narrow emission
lines of the starbursting host.

The
fast velocity measured for the intermediate-width H$\alpha$ component
(\about 6000~\kms) also points towards an extremely energetic
explosion ($> 10^{52}$~ergs) which imparts a faster blast-wave speed
to the postshock material and a higher luminosity from the interaction
than is observed in typical SNe~IIn such as SN~1988Z.  The large
amount of radiated energy can only be achieved if the initial
explosion has a large total radiative and kinetic energy and most of
its kinetic energy is converted into radiation by interaction with a
CSM.  A conventional core-collapse SN explosion can produce at most a
few times $10^{52}$~ergs of energy that can couple to baryonic matter.
The measured kinetic energy of \LMC902 is comparable to this limit
(and similar to GRB-associated SNe), and therefore \LMC902 may have
been produced by an alternative explosion mechanism such as the
pair-instability SN.


The {\it Spitzer} IR photometry shows significant variations in the IR
from 584~days to 683~days after maximum in restframe
3.5, 4.5, and 6.2~\micron. Such a significant change in flux over such
a short time span makes it highly likely that at least some of the
measured infrared flux is caused by the SN. The changes in flux cannot
be explained by a simple change in temperature and/or flux of a
blackbody, but rather requires emission bands of molecules or
dust. We find that the observed increase in the IR flux density at 4.5
and 8.0~\micron\ is consistent with an IR echo of the SN produced by
normal interstellar dust at a distance \about 0.5~pc from the SN in
combination with PAH emission features.

By any measure, SN~2003ma is a remarkable object. Its discovery and
extensive lightcurve coverage were made possible by long-duration
microlensing surveys. Its prompt identification and initial
spectrocopy reveal the power of well-implemented reduction pipelines
and planned, timely coordination of large telescope facilities. The
inclusion of satellite data provided important constraints of the
environment of this object. Ground-based facilities devoted entirely
to deep, time-domain discovery such as the Large Synoptic Survey
Telescope and PanSTARRS are almost certain to reveal many more such
objects and lead us towards even greater understanding of the
progenitors required to produce extremely luminous core-collapse
supernovae like SN~2003ma and illuminate the full range of
observational characteristics for what, for now, remains a small and
elusive set of truly cataclysmic events.


\appendix
\section{Classification of \LMC902}\label{sec:class}

In this section we summarize why we classify \LMC902 as a SN and not
one of these other options.

\subsection{Event in Milky Way or LMC}

From Figure~\ref{fig:positionplot}, we see that \LMC902 is
$0.064\pm0.012$~arcsec from a resolved source in our template images.
In 2008, \LMC902 was much fainter than this source; therefore, our
2008 spectrum is dominated by this source, a starburst galaxy at $z =
0.289$. If the event is in the LMC or Milky Way, then it must be
either a transient or variable source with a quiescent brightness
$\gtrsim 5$~mag fainter than at peak.  This excludes most flaring
variable stars such as M dwarfs and Ae/Be stars since the amplitude of
their flares are generally small compared to the total flux.
Additionally, we would not expect a prolonged, $>5$~year decline from
peak.

Another candidate is a nova, which could flare by $> 5$~mag.  However,
our spectra from 2003 does not show any emission lines at or near zero
redshift, which we would expect since the event is 0.5 magnitude
brighter than the source at that time and therefore should dominate
the spectrum. Furthermore, the light curve is unlike any nova yet
observed.

Considering all this, it is unlikely that \LMC902 occurred in
either the LMC or Milky Way.

\subsection{Supermassive Black Hole Event}

AGN are often variable at various time-scales. Here we discuss if it
is likely that \LMC902 is an AGN.

\begin{itemize}

\item If \LMC902 originates from a black hole in the supposed host galaxy
then the supermassive black hole is offset by 500~pc from the center
of the galaxy.

\item We calculate the weighted average of 42 \mbox{SuperMACHO} \VR\
difference image fluxes before the event (MJD $\le 52965$) spanning 2
years, and find that it is within $1\sigma$ of zero, with a reduced
$\chi^2=1.2$. None of the 42 detections deviates from zero by more
than $3\sigma$, and typical $3\sigma$ lower limits on variability are
24.2 magnitudes. If we split up these 42 detections into three blocks
$52224 \le \mbox{MJD} \le 52291$, $52547 \le \mbox{MJD} \le 52643$,
and $52932 \le \mbox{MJD} \le 52965$, we find again that the average
fluxes are consistent with zero with reduced $\chi^2$ of 1.01, 0.96,
and 1.73, respectively. Therefore the pre-event variability in \VR\ is
fainter than 24.2 magnitude during this 2-year time period.

We perform a similar analysis with the $I$~band light curve and find
that the pre-event OGLE-II ($\mbox{MJD} < 52091$) and OGLE-III ($52091
\le \mbox{MJD} < 52965$) data is consistent with being constant with a
reduced $\chi^2$ of 1.65 and 1.51, respectively. We $3\sigma$-clipped
4\% of the data. When we bin the data (see Section~\ref{sec:optphot}
and bottom panel in Figure~\ref{fig:TDflux}), we find that all average
fluxes are within $3\sigma$ of zero, with reduced $\chi^2$ ranging
from 0.67 to 1.91.

\item Another test for the presence of AGN activity in the host galaxy are
the ratios of the narrow emission lines, which are measured for all
spectra and presented in Table~\ref{tab:ratios2}.  The
[\ion{O}{3}]/H$\beta$ and [\ion{N}{2}]/H$\alpha$, and
[\ion{S}{2}]/H$\alpha$ ratios in the standard diagnostic line ratios
\citep{BPT81, Veilleux87} classify the host galaxy as a star-forming
galaxy.  Although the emission lines may have a contribution from the
transient object, in the post-peak spectra from $2004-2008$ the
[\ion{O}{3}]/H$\beta$, [\ion{N}{2}]/H$\alpha$, and
[\ion{S}{2}]/H$\alpha$ ratios are constant within the errors, which
implies that at late times the narrow-line emission is dominated by
the host galaxy.  

\item We consider the rare case in which an otherwise dormant
supermassive black hole tidally disrupts a star, and emits a flare of
radiation from the accretion of the stellar debris.  This could
explain that the event itself is about 0.5 magnitude brighter than the
host and shows no variability pre-event. These events are typically
bright in the X-rays \citep{Komossa02, Esquej07} and UV
\citep{Gezari06, Gezari08, Gezari09}, however several candidates have
emerged that have also been detected at optical wavelengths
\citep{Gezari08, Gezari09}.  The strong evolution of the foreground
extinction-corrected colors of \LMC902 from $V-I \approx 0$ at peak to
$V-I > 1$ for times later than 1 year after the peak (see
Section~\ref{sec:absmag}), however, are inconsistent with the steady
blue colors expected for the blackbody spectral energy distributions
and effective temperatures associated with tidal disruption events
($T_{\rm BB} > 1\times 10^{4}$ K; \citealt{Gezari09}).
\end{itemize}

The narrow-line ratios, listed in Table \ref{tab:ratios2}, are all
consistent with a star-forming galaxy.  This, in combination with the
lack of variability in the densely sampled \mbox{SuperMACHO} and OGLE
difference imaging data over a baseline of 7 years before the SN,
place strong constraints on the presence of an AGN in the host galaxy.

\subsection{Supernova}

Since \LMC902 is unlikely to be caused by a supermassive black hole or
Galactic or LMC variable, the remaining likely transient phenomenon is
that of a SN.  Its blue continuum in 2003 (see Figures~\ref{fig:spec}
and \ref{fig:BBfit}) and \about 6000~\kms\ FWHM emission lines
consistent with H$\alpha$ in 2004 and 2008 are reminiscent of a SN of
type IIn \citep{Schlegel90}.  This identification is also consistent
with the host galaxy being a star-forming galaxy. In
Sections~\ref{ss:spec_comp} and \ref{ss:phot_comp}, we compare the
spectroscopy and photometry of \LMC902 to other SNe from the
literature, and we find that the most consistent explanation for 
this event is that it is a SN~IIn.

\begin{acknowledgments} 

We thank the referee for many insightful comments that helped to
improve this paper.  The \mbox{SuperMACHO} survey was undertaken under
the auspices of the NOAO Survey Program. We are very grateful for the
support provided to the Survey program from the NOAO and the National
Science Foundation. We are particularly indebted to the scientists and
staff at the Cerro Tololo Inter-American Observatory for their
assistance in helping us carry out the survey.  \mbox{SuperMACHO} is
supported by the STScI grant GO-10583.
AR thanks the NOAO Goldberg Fellowship Program for its support. AG's,
KHC's, MEH's, and SN's work was performed under the auspices of the
U.S.  Department of Energy by Lawrence Livermore National Laboratory
under Contract DE-AC52-07NA27344.  C.~Stubbs thanks the the McDonnell
Foundation for its support through a Centennial Fellowship.
C.~Stubbs, AG, and AR are also grateful for support from Harvard
University. AC and GP acknowledges the support of grant P06-045-F
ICM-MIDEPLAN.  DM, AC, and GP are supported by grants FONDAP CFA
15010003 and Basal CATA 0609.  LM is supported by grant
(CPDR061795/06) from Padova University.  DLW acknowledges financial
support in the form of a Discovery Grant from the Natural Sciences and
Engineering Research Council of Canada (NSERC).  LW acknowledges
support by the EC FR7 grant PERG04-GA-2008-234784.
Based on observations obtained as part of the GS-2003B-Q-12 science
program at the Gemini Observatory, which is operated by the
Association of Universities for Research in Astronomy, Inc., under a
cooperative agreement with the US National Science Foundation on
behalf of the Gemini partnership: the NSF (United States), the Science
and Technology Facilities Council (United Kingdom), the National
Research Council (Canada), CONICYT (Chile), the Australian Research
Council (Australia), Minist\'{e}rio da Ci\^{e}ncia e Tecnologia
(Brazil) and Ministerio de Ciencia, Tecnolog\'{i}a e Innovaci\'{o}n
Productiva (Argentina).  The OGLE project is partially supported by
the Polish MNiSW grant N20303032/4275.

\end{acknowledgments}


\bibliographystyle{fapj}
\bibliography{ms}


\end{document}